\documentclass[a4paper,10pt]{article}
\pdfoutput=1

\newcommand{\be}{\begin{equation}}
\newcommand{\ee}{\end{equation}}

\newcommand{\bea}{\begin{eqnarray}}
\newcommand{\eea}{\end{eqnarray}}

\newcommand{\D}{{\cal D}}
\newcommand{\U}{{\cal U}}

\newcommand{\N}{{\cal N}}
\newcommand{\M}{{\cal M}}

\newcommand{\bbbar}{{b \bar{b}}}

\usepackage{caption}
\usepackage{subcaption}

\makeatletter
\newcommand*{\rom}[1]{\expandafter\@slowromancap\romannumeral #1@}
\makeatother

\usepackage{float}
\usepackage{colortbl}
\usepackage{graphicx}%
\usepackage{multirow}%
\usepackage{amsmath,amssymb,amsfonts}%
\usepackage{amsthm}%
\usepackage{mathrsfs}%
\usepackage[title]{appendix}%
\usepackage{xcolor}%
\usepackage{textcomp}%
\usepackage{manyfoot}%
\usepackage{booktabs}%
\usepackage{algorithm}%
\usepackage{algorithmicx}%
\usepackage{algpseudocode}%
\usepackage{listings}%
\usepackage{bm}
\usepackage{fancybox}

\usepackage[section]{placeins}

\definecolor{Gray}{gray}{0.95}
\definecolor{RGray}{gray}{0.85}
\definecolor{CGray}{gray}{0.92}

\usepackage{multicol}
\definecolor{tit}{rgb}{0.1,0.2,0.4}
\definecolor{blus}{cmyk}{1,1,0,0.6}
\definecolor{verde}{cmyk}{0.92,0,0.59,0.25}

\newcommand*{\myalign}[2]{\multicolumn{1}{#1}{#2}}

\usepackage{
graphicx,
hyperref,
amsmath,
amssymb,
charter,
xcolor,
ifluatex,
longtable,
booktabs,
multirow}
\usepackage{float}
\usepackage[utf8]{inputenc}
\newcommand{\e}[1]{\cdot 10^{#1}}

\usepackage{
graphicx,
hyperref,
amsmath,
amssymb,
charter,
xcolor,
ifluatex,
longtable,
booktabs,
multirow,
bbold}

\usepackage{hyperref}

\usepackage{tipa}
\usepackage{amsmath,amssymb,amsfonts, bm}
\usepackage{dsfont}
\usepackage{epsfig}
\usepackage{graphicx}
\usepackage{slashed}
\usepackage{color}

\usepackage{caption}
\usepackage{subcaption}
\usepackage{slashed}

\usepackage{commath}
\usepackage{calc}

\theoremstyle{thmstyleone}%
\theoremstyle{thmstyletwo}%

\theoremstyle{thmstylethree}%

\begin{document}

\allowdisplaybreaks
\vspace*{-2.5cm}
\begin{flushright}
{\small
IIT-BHU
}
\end{flushright}

\vspace{2cm}

\begin{center}
{\LARGE \bf \color{tit}
 The problem of flavour}\\[1cm]

{\large\bf Gauhar Abbas$^{a}$\footnote{email: gauhar.phy@iitbhu.ac.in}   }  
\\[7mm]

{\large\bf Rathin Adhikari$^{b}$\footnote{email: rathin@ctp-jamia.res.in}   }  \\[7mm]

{\large\bf Eung Jin Chun$^{c}$\footnote{email: ejchun@kias.re.kr}   }  \\[7mm]

{\large\bf Neelam Singh$^{a}$\footnote{email: neelamsingh.rs.phy19@itbhu.ac.in}   }  \\[7mm]

{\it $^a$ } {\em Department of Physics, Indian Institute of Technology (BHU), Varanasi 221005, India}\\[3mm]

{\it $^b$ } {\em Centre for Theoretical Physics, Jamia Millia Islamia (Central University),\\ New Delhi , 110025, India}\\[3mm]

{\it $^c$ } {\em Korea Institute for Advanced Study, Seoul, 02455, Republic of Korea}\\[3mm]

\vspace{1cm}
{\large\bf\color{blus} Abstract}
\begin{quote}
We review the problem of flavour tracing back to the days when the standard model was just coming together. We focus on the recently discussed new solutions of this problem, namely the Froggatt and Nielsen mechanism based on a novel discrete $\mathcal{Z}_{\rm N} \times \mathcal{Z}_{\rm M}$ flavour symmetry,  and the standard hierarchical VEVs model. The standard HVM,  and  the Froggatt and Nielsen mechanism based on the $\mathcal{Z}_{\rm N} \times \mathcal{Z}_{\rm M}$ flavour symmetry,  can be  recovered from a new dark-technicolour paradigm, where the hierarchical VEVs or the flavon VEV  may appear as the chiral multifermion condensates.  In particular, there appears a novel feature that  the solution of the flavour problem based on  the discrete flavour symmetry can provide the so-called flavonic dark matter. This predicts a specific relation between the mass and the symmetry-breaking scale, which can be contrasted with the standard QCD axion. Moreover, a possible direction towards the Grand Unified framework is also discussed.
\end{quote}

\thispagestyle{empty}
\end{center}

\begin{quote}
{\large\noindent\color{blus} 
}

\end{quote}

\newpage
\setcounter{footnote}{0}

\section{Introduction}\label{sec1}
The problem of flavour of the standard model (SM) is one of the most challenging observational puzzles and is defined by the lack of any mechanism to explain the observed  fermionic mass spectrum and mixing, including that of neutrino as well.  This problem is so fascinating that  in his remarkable review `` The Problem of Mass"  in 1977, when the SM  was finally starting to come together, Weinberg wrote, ``Over the
last decade, we have seen a satisfying synthesis of the theories of weak, electromagnetic, and strong interactions, which has provided explanations for many of
the ad-hoc hypotheses that had been previously introduced into particle physics
on chiefly empirical grounds. However, one essential element of this systematic
theory has remained obscure: we must take the masses of the leptons and quarks as
input parameters, without any real idea of where they came from"\cite{Weinberg:1977hb}. Reviews of the flavour problem can be found in references \cite{Raby:1995uv,Peccei:1997mz,Fritzsch:1999ee,Xing:2014sja,Feruglio:2015jfa}.

The earliest attempt to predict the quark-mixing were made by Cabibbo and Franceschi by suggesting that the requirement of vanishing divergent contribution  in the lowest order in weak interactions could produce the correct value of the Cabibbo angle\cite{Cabibbo:1968f}.  This observation was used by  Gatto,  Sartori, and Tonin to relate the Cabibbo angle  to parameters of strong interactions by requiring that   the quadratic divergence of the weak correction does not contribute to the $SU(3)$ and isospin breaking \cite{Gatto:1968ss}.  In the same issue of  Phys. Lett. B,  Cabibbo and Maiani produced the Cabibbo angle by proposing that the weak and electromagnetic  corrections determine the structure of symmetry breaking in the strong   Hamiltonian \cite{Cabibbo:1968vn}.  In 1969,  Oakes predicted the Cabibbo angle by assuming   that chiral $SU(2) \times SU(2)$ breaking in strong interactions and the non-conservation of strangeness in weak interactions can produce the  Cabibbo angle\cite{Oakes:1969vm}. Following this, in 1970, Cabibbo and Maiani again came up with a new model  to predict the Cabibbo angle by relating it to the observed breaking of the $SU(3) \times SU(3)$ symmetry \cite{Cabibbo:1970uc}.

In this paper, after reviewing the main standard solutions to the flavour problem,  we shall discuss some new typical and atypical ideas to address the flavour problem and dark matter together.  For instance, a typical unified solution to the flavour problem and dark matter is the models based on the $\mathcal{Z}_N \times \mathcal{Z}_M$ flavour symmetry \cite{Abbas:2018lga,Abbas:2022zfb,Abbas:2023ion,Abbas:2024dfh}.  We notice that  the  FN mechanism within minimal framework of the SM (non-supersymmetric) based on the $\mathcal{Z}_N \times \mathcal{Z}_M$ flavour symmetry or any discrete symmetry is the first-ever such construction in literature.  This idea replaces the requirement of a gauged symmetry $U(1)$ symmetry,  thus, providing an entirely unique framework with unique signatures at colliders \cite{Abbas:2024dfh}.  The reference  \cite{Abbas:2024dfh}  is a phenomenological guide to flavour models based on the $\mathcal{Z}_N \times \mathcal{Z}_M$ flavour symmetries, and can be used to probe flavor models based on  $\mathcal{Z}_N \times \mathcal{Z}_M$ flavour symmetries at the high-luminosity Large-hadron Collider (HL-LHC), high-energy LHC (HE-LHC)  \cite{FCC:2018bvk}, and 100 TeV collider such as FCC-hh \cite{FCC:2018byv}.  Thus, this review serves a concise introduction to the flavour models based on  $\mathcal{Z}_N \times \mathcal{Z}_M$ flavour symmetries for people working on collider phenomenology.

Further motivation for this review comes  from recent hints of new physics, referred to  as   anomalies,  emerging from different experiments \cite{Crivellin:2023zui}.  The energy scales of these anomalies range from approximately $1$ GeV to the TeV scale probed by the Large Hadron Collider (LHC) \cite{Crivellin:2023zui}.  In particular, a recent review by the ATLAS experiment at the LHC  provides hints of several anomalies spanning from 10 GeV  to the TeV scale  \cite{ATLAS:2024itc}.  For instance, a part of the summary  table  given in reference  \cite{ATLAS:2024itc} is shown in table \ref{tab:anomalies}.  

\begin{figure}
	\centering
	 \includegraphics[width= 0.55\linewidth]{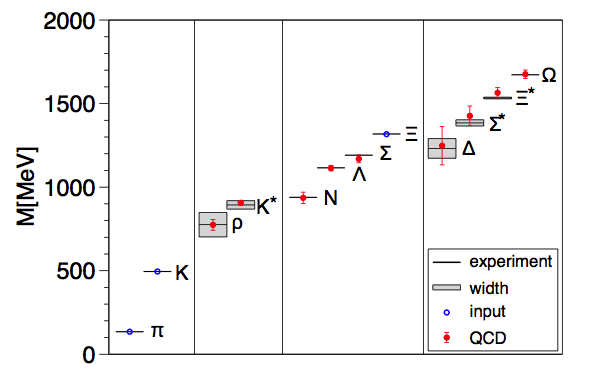}	
\caption{ The light hadron spectrum of QCD. Figure is taken from reference \cite{Battaglieri:2014gca}.  }
\label{hadron}
\end{figure}

We observe from table \ref{tab:anomalies} that  the mass spectrum is quite random,  and apparently is very difficult to fit in any model based on the extended scalar sector of the SM.  We notice that if these anomalies continue to show up in future runs of the LHC, the situation would be  quite similar to the 1950s, when  bubble chambers and spark chambers experiments discovered so many hadrons.  For instance,  see figure \ref{hadron}, taken from reference \cite{Battaglieri:2014gca}.

\begin{table}[h!]
\begin{center}
\resizebox{0.6 \textwidth}{!}{
\begin{tabular}{l|c|c|}
\multirow{2}{*}{\textbf{Decay channel}} &	 \textbf{Production}	&	\multirow{2}{*}{\textbf{Mass GeV}} \\
& \textbf{mode}					&	\\ \hline
$H\to\tau\tau$ &											$b$-associated &		400 	\\			
$H\to\tau\tau$ &											ggF &								400 		 \\
$H\to\mu\mu$ &												$b$-associated &		480 \\
$H\to t\bar{t}$ &											ggF &								800 	 \\
$H\to t\bar{t}/t\bar{q}$ &						qq and qg &					900  \\
$H\to ZZ\to 4\ell /2\ell 2\nu$ &			ggF &								240  \\
$H\to ZZ\to 4\ell /2\ell 2\nu$ &			VBF &								620  \\
$H\to \gamma\gamma$ &									ggF &								684 	 \\
$H\to \gamma\gamma$ &									ggF &								~~~~~95.4  \\
$H\to Z(\ell\ell)\gamma$ &						ggF &								420	 \\
$H\to Z(q\bar{q})\gamma$ &						ggF &								3640 \\
$A\to Zh_{125}(b\bar{b})$ &						ggF &								500  \\
$A\to Zh_{125}(b\bar{b})$ &						$b$-associated &		500  \\
$A\to ZH\to \ell\ell b\bar{b}$ &			ggF	&								610 ($A$), 290 ($H$)  \\
$A\to ZH\to \ell\ell b\bar{b}$ &			$b$-associated &		440 ($A$), 220 ($H$) \\
$A\to ZH\to \ell\ell WW$ &						ggF	&								440 ($A$), 310 ($H$)  \\
$A\to ZH\to \ell\ell t\bar{t}$ &			ggF	&								650 ($A$), 450 ($H$)  \\
$A\to ZH\to Zh_{125}(b\bar{b})h_{125}(b\bar{b})$ &	VH &	420 ($A$), 320 ($H$)   \\
$H^+\to cb$ &													$t\bar{t}$ decay &	130  \\
$H^+\to Wa(\mu\mu )$ &								$t\bar{t}$ decay &	120--160 ($H^+$), 27 ($a$)  \\
$H^{++}\to WW$ &											VBF &								450 	 \\
$H\to h_{125}h_{125}\to 4b$ &					ggF &								1100 \\
$H\to h_{125}h_{125}\to 4b$ &					VBF &								550 	 \\
$H\to h_{125}h_{125}\to b\bar{b}\tau\tau$ &	ggF &					1000~~  \\
$H\to h_{125}h_{125} \, \text{combination}$ &	ggF &				1100~~ \\
$X\to Sh_{125}\to \bbbar\gamma\gamma$ &	ggF &							575 ($X$), 200 ($S$)  \\
$h_{125}\to Z_dZ_d \to 4\ell$	&				ggF &								~~28  \\
$h_{125}\to ZZ_d \to 4\ell$ &					ggF &								~~39 \\
$h_{125}\to aa\to b\bar{b}\mu\mu$ &		ggF, VBF, VH &			~~52 \\
$h_{125}\to aa\to 4\gamma$ &					ggF &								10--25~~~~  \\
$h_{125}\to e\tau$ and $h_{125}\to \mu\tau$ &		ggF, VBF, VH &	125~~ 
 \end{tabular}
}
\caption{  The anomalies reported by the ATLAS experiment.  This table is adopted from reference \cite{ATLAS:2024itc}.}
\label{tab:anomalies}
\end{center}
\end{table}

The first guess in that scenario would be to assume that these anomalies  are not fundamental particles, and rather bound states of a new strong force.  Along this line of argument, we shall discuss new models of flavour based VEVs hierarchy \cite{Abbas:2017vws,Abbas:2023dpf}.  In particular,  the Standard Hierarchical VEVs Model (SHVM) \cite{Abbas:2023dpf}, may have potential to address the hierarchical anomalies given in table \ref{tab:anomalies}.   Remarkably, an  ultra-violet (UV) completion of these models  lies in a strong dark technicolour paradigm based on $\mathcal{G} \equiv SU(\rm N_{\rm TC}) \times SU(\rm N_{\rm DTC}) \times SU(\rm{N}_{\rm F})$ symmetry, where TC stands for technicolour, which may be a conformal strong dynamics,  DTC for dark-technicolour, and F represents strong dynamics of dark-QCD of vector-like fermions \cite{Abbas:2020frs,Abbas:2023bmm}. 

In this framework, the hierarchical VEVs  are chiral multi-fermion bound states of the $SU(\rm N_{\rm DTC})$ dynamics. Moreover, $ SU(\rm{N}_{\rm F})$ is a QCD-like dynamics.  Therefore, its spectrum can be predicted by rescaling the spectrum of QCD. Thus, it may explain several anomalies given in table \ref{tab:anomalies},  and early investigation shows that the 95.4 GeV excess can easily accounted in the SHVM framework \cite{Abbas:2024jut}.  Furthermore, the  spectrum of $SU(\rm N_{\rm DTC})$ dynamics, a part of which exhibits itself as the gauge singlet scalar fields in the SHVM, may account for several anomalous data given in table \ref{tab:anomalies}. An additional purpose of this review is to introduce the community to the unconventional  approach of the SHVM, and its ultra-violet (UV) completion based on the dark-technicolour paradigm.

Apart from above discussion, we have presented unified solutions of the flavour and dark matter in this review.  The models based on $\mathcal{Z}_N \times \mathcal{Z}_M$ flavour symmetries contain flavonic dark matter.  On the other side, the SHVM may also provide a new kind of ``possible" dark matter particle.  Thus, apart from collider researchers,  the review may also be interesting for the people working on dark matter searches, and for people working on theoretical models of scalar dark matter.

The plan of the review is as follows: In section \ref{sec2}, we review the flavour models based on continuous and discrete symmetries,  GUT theories, compositeness hypothesis, and extra dimensions. Section \ref{sec3} has the discussion of the flavour bounds on two prototype forms of the $\mathcal{Z}_{\rm N} \times \mathcal{Z}_{\rm M}$ flavour symmetry.  In section \ref{sec4}, we show that the $\mathcal{Z}_{\rm N} \times \mathcal{Z}_{\rm M}$ flavour symmetry can provide a unique dark matter particle named flavonic dark matter.  We discuss models based on the VEVs hierarchy in section \ref{sec:HVM}.  A UV completion of the models based on the VEVs hierarchy and the $\mathcal{Z}_{\rm N} \times \mathcal{Z}_{\rm M}$ flavour symmetry, which employs a dark technicolour symmetry, is discussed in section \ref{UV}. We present  an outline of an extended  and dark-extended technicolour scenario for the dark-technicolour paradigm in section \ref{ETC}. We summarize in section \ref{sec7}.

\section{Solutions of the flavour problem}
\label{sec2}
Weinberg was the first to pay attention to the problem of the fermionic mass spectrum in  theories with spontaneously broken gauge symmetries\cite{Weinberg:1971nd}. He  incorporated the weak and electromagnetic interactions  into a parity-conserving  $SU(3) \times SU(3)$ parity invariant gauge theory, which is spontaneously broken to the $SU(2)_L \times U(1)_Y$ model. In this model, the leptons $\mu^+$, $e$, and $\nu$ form a Konopinski-Mahmoud triplet, and a scalar field $\phi$ is added, transforming as $(3,\bar{3})$ representation and coupled to the leptonic triplet in a $SU(3) \times SU(3)$ invariant manner.  The scalar field $\phi$ acquires the VEVs only for the diagonal elements due to the charge conservation.  It turns out that due to the smallness of leptonic masses, only non-zero VEV is $\langle \phi_{11} \rangle$.  The scalar field components $(\phi_{11}, \phi_{21})$ form a scalar Higgs doublet which further breaks the SM $SU(2)_L \times U(1)_Y$ symmetry spontaneously.  The mass of the electron is predicted to be  $\alpha m_\mu$ from the interactions of leptonic triplet and the scalar field $\phi$.

In fact, Weinberg was so obsessed with the flavour problem that it became the only mystery that he wanted to see being solved in his lifetime\cite{Abbas:2017vws}.  He recalled this in his interview with CERN Courier, ``It was the worst summer of my life! I mean, obviously, there are broader questions, such as: Why is there something rather than nothing? But if you ask for a very specific question, that’s the one. And I’m no closer now to answering it than I was in the summer of 1972”\cite{cern_courier}.

However, even the worst summer of his life gave birth to  the first pioneer and standard  paradigm for the solution of the flavour problem  by exploiting the fact that the representation content of the fields and symmetry can be used to eliminate the masses of light fermions at tree-level, and as a consequence, the masses of light fermions can  be calculated as finite higher-order effects \cite{Weinberg:1972ws}.   The main obstacle in front of Weinberg was that, a symmetry, which forbids some mass or mass differences at zeroth order, also does not allow the mass or mass difference nonzero in all higher orders.  However, the renormalizability of  spontaneously broken gauge symmetries, such as the SM,  provides a way out of this impasse.  In a renormalizable spontaneously broken gauge theory, if the zeroth-order contribution to a  mass or mass difference vanishes for a given set of fields in some representation of the gauge symmetry, then the higher order contribution to the mass or mass difference must be finite.  Assuming an extreme scenario, Weinberg argued that if there are no scalar fields belonging to the irreducible representation of the symmetry at hand, the Yukawa couplings are completely forbidden.  However, the VEVs of the scalar fields (for example, the Higgs field in the SM) contribute to masses of the vector bosons of the corresponding gauge symmetry.  The higher order corrections introduced by the vector gauge bosons can produce the finite fermion masses.  This powerful idea was immediately used by Georgi and Glashow to explore the fermionic mass patterns using the different representations of  scalar multiplets in Abelian and non-Abelian models\cite{Georgi:1972mc,Georgi:1972hy}.  

One of the most important ideas for solving the flavour problem was put forward by the FN in 1979\cite{Froggatt:1978nt}. FN were  inspired by the idea of Weinberg of generating fermionic masses through symmetry and the representation of scalar fields. In the FN mechanism,  an Abelian flavour symmetry $U(1)$ and a gauge singlet scalar are used to produce the masses of fermions through the effective operators having the following structure,
\begin{equation}
\mathcal{O} = y (\dfrac{ \chi }{\Lambda})^{(\theta_i + \theta_j)} \bar{\psi} \varphi \psi,
\end{equation}
where $y$ is the coupling constant, $\chi$ is the gauge singlet scalar field known as flavon, $\varphi$ represents the SM Higgs field, $\theta_i$ and $\theta_j$ are the charges of the fermions $\psi_i^c$ and $\psi_j$ under  $U(1)$ symmetry, and $\Lambda$ is the scale at which these interactions become part of a renormalized underlying theory.

The solutions to the flavour problem can be divided into two classes,  bottom-up and top-down.  Bottom-up solutions are designed for relatively higher and accessible energies, and they may be a subset of a larger underlying theory.  On the other side,  the top-down scenarios are motivated by GUT and other problems, such as the hierarchy problem of the SM.    

Bottom-up solutions are mainly based on continuous and discrete symmetries \cite{Froggatt:1978nt,flavor_symm1,Chun:1996xv,flavor_symm2,flavor_symm3,Davidson:1983fy,Davidson:1987tr,Berezhiani:1990wn,Berezhiani:1990jj,Berezhiani:1989fp,Sakharov:1994pr,Hinata:2020cdt,Abbas:2018lga,Abbas:2022zfb,Higaki:2019ojq,Abbas:2024wzp,Abbas:2024dfh}.
Soon after Weinberg's 1972 paper, serious efforts were being made to address the flavour problem in parts.  For instance, Georgi and Pais discussed 
calculability conditions  for local gauge theories with spontaneous symmetry breaking and applied them to create a mechanism to compute the Cabibbo angle  in models based on $SU(2) \times U(1)  \times U(1) $ and $O(4) \times U(1) $ symmetries \cite{Georgi:1974yw}.

Georgi and Pais used  the fact that the counter-terms required for renormalizability of the SM possess the symmetries of the Lagrangian before spontaneous breaking of the symmetry. This produces  non-trivial relations  among the counter-terms, resulting  in zeroth-order relations among masses and couplings. This important result can be used to create non-trivial relations between the Cabibbo angle and the masses of quarks.  For example, $\cos^2 \theta_W = M_W^2/M_Z^2$  is an example of the zeroth-order relation in the SM.  For instance, in the case of $SU(2) \times U(1)  \times U(1) $ model with two quark doublets, the Higgs multiplets required to generate leptonic masses are different from those which create quark masses.  For quark masses, the model contains three Higgs doublets $\phi$, $\chi$, and $\eta$ of the type $(\frac{1}{2}^{(0,-1/2)})$.  Now, by imposing $\mathcal{Z}_2$-type discrete symmetries, one can derive the Cabibbo angle in terms of the masses of the quarks of two doublets \cite{Georgi:1974yw}.

The first serious attempt to explain the flavour mixing in a gauge theory with spontaneous symmetry breaking was made by  Rujula, Georgi, and  Glashow (RGG) in an $SU(2)_L \times SU(2)_R \times U(1) $ model where quark mixing is produced through radiative corrections in the spirit of Weinberg's 1972 model\cite{DeRujula:1977dmn}.  RGG used a discrete symmetry to forbid the off-diagonal terms in fermionic mass matrices such that the charged weak current is flavour diagonal.  The flavour mixing is obtained by the soft breakdown of the discrete symmetry in the mass terms of the Higgs field resulting in a calculable Cabibbo angle in terms of a small soft breaking parameter.  Shortly after this,  Fritzsch calculated  the Cabibbo angle in terms of the quark mass ratios in an $SU(2)_L \times SU(2)_R \times U(1) $ model and a discrete symmetry\cite{Fritzsch:1977za}.

In 1979, Wilczek and  Zee introduced the idea of gauged horizontal continuous symmetries, which can determine the weak mixing angles  in terms of quark masses\cite{Wilczek:1978xi}.  The idea was to introduce a gauge symmetry along the horizontal direction, which acts among the left-handed doublets and among the right-handed singlets.  This is implemented by extending the SM by the continuous symmetry $SU(2)_H$, under which the vertical fermionic doublets and singlet fermions  of the SM transform  as triplets of the  $SU(2)_H$.  By adding two scalar fields, which are behaving as tensors and vectors under the  $SU(2)_H$, one can obtain the Cabibbo angle in terms of masses of $d$, $s$, $u$, and $c$ quarks.

The FN mechanism was published in the same year. The models based on the FN mechanism with  continuous symmetries are discussed in the references \cite{Leurer:1992wg,Leurer:1993gy,Ibanez:1994ig,Binetruy:1994ru,Dudas:1995yu,Grossman:1995hk,Nir:1995bu,Barbieri:1995uv,Binetruy:1996xk,Barbieri:1996ww,Choi:1996se,Nelson:1997bt,Buchmuller:1998zf,Babu:1999me,King:2001uz}.  In the framework of the SM, considering only one Higgs doublet with a $U(1)$ flavour symmetry, the flavour problem has been addressed in \cite{Babu:1999me}.  Flavour symmetry is spontaneously broken  at energy scale $M$ around the TeV scale. The effective theory below energy scale $M$ corresponds to the SM with one Higgs doublet but with non–renormalizable terms in the Higgs Yukawa
couplings. Only the top quark has renormalizable Yukawa couplings. The couplings of other fermions are suppressed by different powers of $(H^\dag H/M^2)$, where $H$ is the Standard Model Higgs doublet. This suppression factor provides the small expansion parameter $\epsilon$, and is subjected to the flavour symmetry, resulting in different powers of $\epsilon$  in  the fermionic mass matrices.  However, for neutrinos whose masses are expected to be in the eV scale, seesaw mechanism has been considered in the work of E. Ma \cite{ma}. In this work, in addition to SM particles, heavy right-handed neutrinos with TeV scale mass has been considered as the new physics scale. In the scalar sector, an additional scalar doublet is also considered.  A $U(1)$ flavour symmetry with lepton number has been considered in such a way that the heavy fermionic singlets interact with the SM leptons only through the additional scalar doublet but not through the SM higgs doublet. As the VEV of the additional doublet is in the MeV scale, the Dirac mass term in the seesaw mass matrix is much lower, and appropriate light neutrino masses are naturally obtained with TeV scale mass of heavy right-handed neutrinos. 

In another work \cite{raar} in obtaining small neutrino masses, certain conditions of Dirac and Majorana masses have been found under which Type-\rom{1} seesaw mechanism can lead to different number of massless neutrinos. Such conditions are found to be related with some underlying $U(1)$ symmetry. In such cases, higher-order corrections at one or two loop level provide the appropriate light neutrino mass with seesaw scale in the TeV range.  

Flavour deconstruction is another interesting approach to address the flavour problem \cite{Bordone:2017bld}-\cite{Greljo:2024ovt}. In the flavour deconstruction framework, the  hierarchy of the fermionic masses  is obtained through a successive hierarchical spontaneous symmetry breaking (SSB) of the gauge symmetries.  For instance, we could have a gauge group $G_1 \times G_2 \times  G_3$, where each $G_i$ represents a gauge symmetry for each fermionic family of the SM.  Now, a chain of SSBs along the track  $G_1 \times G_2 \times  G_3 \rightarrow  G_{12} \times G_3 \rightarrow  G$ facilitated by  the VEVs of some scalar fields,  provides the masses to each generation of fermions.

An alternative way to address the flavour problem is through discrete  symmetries.  The first attempt in this direction was made by Wilczek and Zee (WZ) in 1977 by predicting the relation $\tan^2 \theta_c = m_d/m_s $\cite{Wilczek:1977uh}.  WZ employed the gauge group $SU(2)_L \times SU(2)_R \times U(1) $ with four quark doublets with two bi-doublets followed by the imposition of a discrete symmetry.  The discrete symmetry allows a particular mass matrix, which produces $\tan^2 \theta_c = m_d/m_s $. The idea was further explored in references \cite{Branco:1978qj,Pakvasa:1977in,Wyler:1979fe}. 

The masses of fermions are generated through the radiative corrections using a   horizontal  non-Abelian discrete symmetry $S_3$  in  reference  \cite{Babu:1990fr}. In this model, the non-Abelian discrete symmetry $S_3$  allows only  top and bottom quark masses at tree-level.  The masses of charm, strange quarks and $\tau$-lepton originate at the one-loop level.  The masses of $u$, $d$ quarks, and $\mu$-lepton arise at the two-loop level, and the mass of  electron at the three-loop level.

A minimal symmetry $\mathcal{Z}_2 \times \mathcal{Z}_5$ can solve the flavour problem by implementing the FN mechanism\cite{Abbas:2018lga}.  This symmetry is a prototype of the general $\mathcal{Z}_N \times \mathcal{Z}_M$ flavour symmetry\cite{Abbas:2018lga,Abbas:2022zfb,Abbas:2023ion}.  One can also use a large $Z_N$ symmetry to construct the FN mechanism in a SUSY framework\cite{Higaki:2019ojq}.   Discrete symmetries are also extensively used in neutrino model building\cite{Altarelli:2010gt,King:2013eh,Petcov:2017ggy,Valle}.  Phenomenological investigations of the FN mechanism using continuous and discrete symmetries  are performed in references \cite{Dorsner:2002wi,Giudice:2008uua,Tsumura:2009yf,Berger:2014gga,Diaz-Cruz:2014pla,Huitu:2016pwk,Arroyo-Urena:2022oft,Bauer:2015kzy,Bauer:2016rxs,Abbas:2022zfb}.  

Recently, modular symmetries are also used to address the flavour problem \cite{Feruglio:2017spp}.  A modular symmetry is created by a geometry of a two-dimensional torus $T^2$.  In models based on modular symmetries, the modular invariance acts like a flavour symmetry.  The chiral multiplets reside in unitary representations of the modular group, the Yukawa couplings are the modular forms of level $N$.  In this setup, the flavon field is not an essential requirement, and  the modular invariance can be broken by the VEV of the modulus.  For a review of the modular symmetries, see reference \cite{Kobayashi:2023zzc}.

Top-down solutions to the flavour problem, in particular, involve GUT theories based on Pati-Salam and  $SU(5)$ unification \cite{Pati:1974yy,Georgi:1974sy}. One of the earliest attempts to provide  an explanation to the fermion masses was using $SU(N)$ GUT with $N>5$, which are broken to $SU(5)$ GUT \cite{Barr:1979xt}.  In this approach, fermions are accommodated in a so-called primitive representation, which is left-handed and totally anti-symmetric $SU(N)$ tensor.  Moreover, this representation is anomaly-free and produces three fermionic families of the SM with  the existence of a hierarchy of masses among light fermions. To achieve this, replication of the primitive representation is also required.  This is followed by solutions based on $SO(10)$ GUT\cite{Barr:1981wv,Davidson:1983fy}.  A different solution is constructed using $O(10) \times U(1)_F$ GUT in reference \cite{Dimopoulos:1983rz}.

The renormalization-group equations can also provide the evolution of Yukawa coupling from a large scale, such as the GUT scale, through fixed-points leading to predictions of fermions masses \cite{Hill:1980sq}. In this approach, the low-energy structure of the renormalization-group equations determines the SM  fermion masses
through the evolution from a scale $M_X$ to $m_f$.  This is obtained by a solution of   the renormalization-group equations, referred to as intermediate-fixed-point,  that is relevant at $\mu^2=m_f^2$.  More recent studies are in references \cite{Ananthanarayan:1991xp,Hempfling:1993kv,Hall:1993gn,Anderson:1993fe,Murayama:1995fn,Rattazzi:1995gk,Arkani-Hamed:1995nqz,Baer:2001yy,Blazek:2002ta,Ross:2002fb,Joshipura:2012sr,King:2013hoa}.

Another class of  top-down solutions to the flavour problem is based on extra-dimensions models\cite{Arkani-Hamed:1998jmv,Randall:1999ee}.  For instance, a five-dimensional model where the bulk is a slice of AdS, and warped extra dimensions can provide a solution to the flavour problem \cite{Gherghetta:2000qt,Blanke:2008zb,Fuentes-Martin:2022xnb}.  For instance, reference \cite{Gherghetta:2000qt} allows the SM fermion fields to propagate in the $\rm AdS_5$ bulk,  and localized Higgs field originates from a Kaluza-Klein excitation  on the TeV-brane. This results in a warp factor similar to that is used to provide a solution of the  gauge hierarchy problem by creating the TeV scale from the Planck scale.  This setup produces the four-dimensional Yukawa couplings  \cite{Gherghetta:2000qt},

\begin{equation}
\label{yc}
    Y_{ij}= \frac{Y_{ij}^{(5)}k}{N_{iL} N_{jR}} 
    e^{(1-c_{iL}-c_{jR})\pi kR}~,
\end{equation}
where 
\begin{equation}
\label{norm}
    \frac{1}{N_{iL}^2} \equiv \frac{1/2-c_{iL}}{e^{(1-2c_{iL})\pi kR}-1}~,
\end{equation}

We can generate  exponentially small Yukawa couplings  for values of $c_{iL}$ and $c_{jR}$ slightly larger than
$1/2$  \cite{Gherghetta:2000qt}.  This results in an explanation for the fermion mass hierarchy through the metric warp factor.  Moreover, the issue of neutrino masses can also be addressed in an extra-dimensional framework without invoking a see-saw mechanism\cite{Grossman:1999ra}.  Furthermore, there are Randall-Sundrum-type models, which are also capable of addressing the flavour problem\cite{Casagrande:2008hr,Bauer:2009cf}.

The models based on dynamical symmetry breaking\cite{Weinberg:1975gm,Susskind:1978ms} are also very attractive scenarios for solving the flavour problem.  For instance, the idea of Pati and Salam \cite{Pati:1975md} that the SM fermions may be the bound states of more fundamental PRE-entities can provide a solution to the flavour problem\cite{Kaplan:1991dc}.  In this model, the spontaneous breaking of the $SU(2)_L \times U(1)_Y$ of the SM is performed by a technicolour dynamics.  Unlike conventional technicolour models,  this is achieved by not coupling the SM fermions directly to the order parameter of the $SU(2)_L \times U(1)_Y$ breaking.  Instead, the SM fermions mix with technibaryons.  The diagonalization of the fermion/technibaryon mass matrix  provides the masses of the SM fermions.  Thus, in this model, the SM fermions are semi-composite states, having an admixture of technicoloured states.  

The assumption of composite fermions and an approximate $U(2)^3$ symmetry can also provide a solution, for instance, see reference \cite{Barbieri:2012uh} and references therein.  A dark technicolour symmetry can be used to solve the flavour problem either through the VEVs hierarchy \cite{Abbas:2017vws,Abbas:2020frs,Abbas:2023dpf,Abbas:2023bmm}. This will be discussed in section \ref{UV} of this review.

\section{Models based on the $\mathcal{Z}_{\rm N} \times \mathcal{Z}_{\rm M}$ flavour symmetry}
\label{sec3}
In this section,  we discuss a solution to the flavour problem  based on the $\mathcal{Z}_{\rm N} \times \mathcal{Z}_{\rm M}$  flavour symmetry\cite{Abbas:2018lga}, which  determines the flavour structure, including masses and mixing,  of the SM through the FN mechanism.    In literature, the conventional FN mechanism is based on a $U(1)$ symmetry, which could be a global or a gauged symmetry.  If this symmetry is a global symmetry, we expect a massless Goldstone boson, which is the result of the spontaneous breaking of this symmetry.  If this symmetry is gauged, a massive vector boson will play a non-trivial role in its phenomenology.

The  $\mathcal{Z}_{\rm N} \times \mathcal{Z}_{\rm M}$  flavour symmetry could be a consequence of the spontaneous breaking  of a $U(1) \times U(1)$ symmetry.  The $U(1) \times U(1)$ symmetry could be a global or a local symmetry.  It could be that only one of the $U(1)$ factors is local in  $U(1) \times U(1)$ symmetry.   We must note that what keeps apart the $\mathcal{Z}_{\rm N} \times \mathcal{Z}_{\rm M}$  flavour symmetry from the conventional FN models based on a $U(1)$ symmetry is that, in addition to  a $U(1) \times U(1)$ symmetry, the $\mathcal{Z}_{\rm N} \times \mathcal{Z}_{\rm M}$  flavour symmetry can be originated from  a global  $U(1)_A \times U(1)_A$ symmetry.  This is discussed in section \ref{UV}.

The following Lagrangian  provides masses to the charged fermions of the SM through the  $\mathcal{Z}_{\rm N} \times \mathcal{Z}_{\rm M}$  flavour symmetry,
\bea
\label{mass1}
-{\mathcal{L}}_{\rm Yukawa} &=&    \left[  \dfrac{ \chi(\chi^{\dagger})}{\Lambda} \right]^{n_{ij}^u}     y_{ij}^u \bar{ \psi}_{L_i}^q  \tilde{\varphi} \psi_{R_j}^{u}  +   \left[  \dfrac{ \chi(\chi^{\dagger})}{\Lambda} \right]^{n_{ij}^d}      y_{ij}^d \bar{ \psi}_{L_i}^q  \varphi \psi_{R_j}^{d}  
+   \left[  \dfrac{ \chi(\chi^{\dagger})}{\Lambda} \right]^{n_{ij}^\ell}        y_{ij}^\ell \bar{ \psi}_{L_i}^\ell  \varphi \psi_{R_j}^{\ell} 
+  {\rm H.c.}, \\ \nonumber
&=&  Y^u_{ij} \bar{ \psi}_{L_i}^q  \tilde{\varphi} \psi_{R_j}^{u}
+ Y^d_{ij} \bar{ \psi}_{L_i}^q  \varphi \psi_{R_j}^{d}
+ Y^\ell_{ij} \bar{ \psi}_{L_i}^\ell  \varphi \psi_{R_j}^{\ell}   + \text{H.c.}, 
\eea
where $\chi$ or $\chi^\dagger$ may be used in the numerator, $i$ and $j$   are family indices, $ \psi_{L}^q,  \psi_{L}^\ell  $ denote the  quark and leptonic doublets, $ \psi_{R}^u,  \psi_{R}^d, \psi_{R}^\ell$ represent right-handed up, down-type  quarks and  leptons, $\varphi$ and $ \tilde{\varphi}= -i \sigma_2 \varphi^* $  are the SM Higgs field and its conjugate and $\sigma_2$ denote the second Pauli matrix. The couplings $Y_{ij}$  are the effective Yukawa couplings given by, $Y_{ij} = y_{ij} \epsilon^{n_{ij}}$, where $   \dfrac{\langle \chi \rangle} { \Lambda} = \dfrac{f}{\sqrt{2} \Lambda}= \epsilon <1$.

We follow a simple rule, named the principle of minimum suppression (PMS), to determine  a minimal form of the  $\mathcal{Z}_{\rm N} \times \mathcal{Z}_{\rm M}$  flavour symmetry \cite{Abbas:2022zfb}.  According to this, there is a minimum suppression in the Lagrangian given in equation \ref{mass1}.  Thus, the top quark mass arises from the SM tree level Yukawa operator.  This is followed by the bottom quark having mass from the operator of the order  $y \epsilon$, the operator with suppression of the order  $y \epsilon^2$ provides mass to the charm quark, the strange quark acquires the mass from the operator of the order  $y \epsilon^3$, followed by the  up  and down quarks having the contribution from the order at least of the order $y \epsilon^4$.  Another constraint on the minimal form of the  $\mathcal{Z}_{\rm N} \times \mathcal{Z}_{\rm M}$  flavour symmetry comes from the demand of neutrino masses and mixing.   Using these arguments,  it can be shown that the minimal symmetry, capable of explaining the flavour problem, should be $\mathcal{Z}_{\rm 2} \times \mathcal{Z}_{\rm 5}$\cite{Abbas:2022zfb}.

The two simple examples of the $\mathcal{Z}_{\rm N} \times \mathcal{Z}_{\rm M}$  are given in table \ref{tab_charges} for $\rm N = 2$ and $\rm M=5,9$, where the charge assignments of the various fields under these symmetries are shown.  
  \begin{table}[h]
  \setlength{\tabcolsep}{6pt} 
\renewcommand{\arraystretch}{1.4} 
\begin{center}
\begin{tabular}{|c|c|c|c|c|c||c|c|c|c|c|c|}
  \hline
  Fields             &        $\mathcal{Z}_2$                    & $\mathcal{Z}_5$ & Fields             &        $\mathcal{Z}_2$                    & $\mathcal{Z}_5$ & Fields             &        $\mathcal{Z}_2$                    & $\mathcal{Z}_9$ & Fields             &        $\mathcal{Z}_2$                    & $\mathcal{Z}_9$        \\
  \hline
  $u_{R}, c_{R}, t_{R}$                 &   +  & $ \omega^2$      & $\psi_{L_2}^q$,  $\psi_{L_2}^\ell$                   &   +  &     $\omega^4 $       &  $u_{R}, t_{R}$                 &   +  & $ 1$       & $\psi_{L_1}^q$,  $\psi_{L_1}^\ell$                   &   +  &    $\omega $             \\
   $d_{R},  s_{R}, b_{R}$                 &   -  &     $\omega $           &  $\psi_{L_3}^q$,  $\psi_{L_3}^\ell$                &   +  &      $ \omega^2 $      &  $c_{R}$                 &   +  & $ \omega^4$   &  $\psi_{L_2}^q$,  $\psi_{L_2}^\ell$                  &   +  &     $\omega^8 $            \\
  $ e_R, \mu_R, \tau_R$                 &   -  &     $\omega $              &  $\chi$                        & -  &       $ \omega$      & $d_{R},  s_{R},  b_{R}$                &   -  &     $\omega^3 $    &  $\psi_{L_3}^q$,    $\varphi$                   &   +  &      $ 1 $              \\
  $ \nu_{e_R},   \nu_{\mu_R}, \nu_{\tau_R} $                 &   -  &     $\omega^3 $                              &  $\varphi$              &   +        &     1 & $ e_R, \mu_R, \tau_R $                 &   -  &     $\omega^3 $ & $\psi_{L_3}^\ell$                 &   +  &      $ \omega^6 $     \\
   $\psi_{L_1}^q$,    $\psi_{L_1}^\ell$                   &   +  &    $\omega $                          &              &           &  &  $ \nu_{e_R},   \nu_{\mu_R},  \nu_{\tau_R}  $                 &   -  &     $\omega^7 $ & $\chi$                        & -  &       $ \omega$     \\
  \hline
     \end{tabular}
\end{center}
\caption{Charges of the SM and flavon fields under $\mathcal{Z}_2$, $\mathcal{Z}_5$, and $\mathcal{Z}_9$  symmetries,  where $\omega$ is the fifth or ninth root of unity.  }
 \label{tab_charges}
\end{table}

The  $\mathcal{Z}_{\rm 2} \times \mathcal{Z}_{\rm 5} $ flavour symmetry creates the following   Lagrangian for the charged fermions,
\bea
\label{massz5}
-{\mathcal{L}}_{\rm Yukawa} &=&    \left(  \dfrac{ \chi}{\Lambda} \right)^{4}  y_{11}^u \bar{ \psi}_{L_1}^q \tilde{\varphi} u_{R}+  \left(  \dfrac{ \chi}{\Lambda} \right)^{4}  y_{12}^u \bar{ \psi}_{L_1}^q \tilde{\varphi} c_{R} +  \left(  \dfrac{ \chi}{\Lambda} \right)^{4}  y_{13}^u \bar{ \psi}_{L_1}^q \tilde{\varphi}  t_{R} +  \left(  \dfrac{ \chi}{\Lambda} \right)^{2}  y_{21}^u \bar{ \psi}_{L_2}^q \tilde{\varphi} u_{R}\nonumber \\
&+& \left(  \dfrac{ \chi}{\Lambda} \right)^{2}  y_{22}^u \bar{ \psi}_{L_2}^q \tilde{\varphi} c_{R} + \left(  \dfrac{ \chi}{\Lambda} \right)^{2}  y_{23}^u \bar{ \psi}_{L_2}^q \tilde{\varphi} t_{R}+   y_{31}^u \bar{ \psi}_{L_3}^q \tilde{\varphi} u_{R} +   y_{32}^u \bar{ \psi}_{L_3}^q \tilde{\varphi} c_{R} +   y_{33}^u \bar{ \psi}_{L_3}^q \tilde{\varphi} t_{R} \nonumber \\
&+&  
 \left(  \dfrac{ \chi}{\Lambda} \right)^{5} y_{11}^d \bar{ \psi}_{L_1}^q  \varphi d_{R} + \left(  \dfrac{ \chi}{\Lambda} \right)^{5} y_{12}^d \bar{ \psi}_{L_1}^q  \varphi s_{R} + \left(  \dfrac{ \chi}{\Lambda} \right)^{5} y_{13}^d \bar{ \psi}_{L_1}^q  \varphi b_{R} + \left(  \dfrac{ \chi}{\Lambda} \right)^{3} y_{21}^d \bar{ \psi}_{L_2}^q  \varphi d_{R} \nonumber \\
&+& \left(  \dfrac{ \chi}{\Lambda} \right)^{3} y_{22}^d \bar{ \psi}_{L_2}^q  \varphi s_{R} + \left(  \dfrac{ \chi}{\Lambda} \right)^{3} y_{23}^d \bar{ \psi}_{L_2}^q  \varphi b_{R} + \left(  \dfrac{ \chi}{\Lambda} \right) y_{31}^d \bar{ \psi}_{L_3}^q  \varphi d_{R}+ \left(  \dfrac{ \chi}{\Lambda} \right) y_{32}^d \bar{ \psi}_{L_3}^q  \varphi s_{R}   \nonumber \\ 
&+& \left(  \dfrac{ \chi}{\Lambda} \right) y_{33}^d \bar{ \psi}_{L_3}^q  \varphi b_{R} + 
\left(  \dfrac{ \chi}{\Lambda} \right)^{5} y_{11}^\ell \bar{ \psi}_{L_1}^\ell  \varphi e_{R} + \left(  \dfrac{ \chi}{\Lambda} \right)^{5} y_{12}^\ell \bar{ \psi}_{L_1}^\ell  \varphi \mu_{R} + \left(  \dfrac{ \chi}{\Lambda} \right)^{5} y_{13}^\ell \bar{ \psi}_{L_1}^\ell  \varphi \tau_{R}  \nonumber \\
&+& \left(  \dfrac{ \chi}{\Lambda} \right)^{3} y_{21}^\ell \bar{ \psi}_{L_2}^\ell  \varphi e_{R} + \left(  \dfrac{ \chi}{\Lambda} \right)^{3} y_{22}^\ell \bar{ \psi}_{L_2}^\ell  \varphi \mu_{R} + \left(  \dfrac{ \chi}{\Lambda} \right)^{3} y_{23}^\ell \bar{ \psi}_{L_2}^\ell  \varphi \tau_{R} +  \left(  \dfrac{ \chi}{\Lambda} \right) y_{31}^\ell \bar{ \psi}_{L_3}^\ell  \varphi e_{R} \nonumber \\
&+& \left(  \dfrac{ \chi}{\Lambda} \right) y_{32}^\ell \bar{ \psi}_{L_3}^\ell  \varphi \mu_{R} \nonumber + \left(  \dfrac{ \chi}{\Lambda} \right) y_{33}^\ell \bar{ \psi}_{L_3}^\ell  \varphi \tau_{R} 
 + \text{H.c.}
\eea

The mass matrices for  up- and down-type quarks and charged leptons  can be written as,
\begin{equation}
\M_u = \dfrac{v}{\sqrt{2}}
\begin{pmatrix}
y_{11}^u  \epsilon^4 &  y_{12}^u \epsilon^4  & y_{13}^u \epsilon^4    \\
y_{21}^u  \epsilon^2    & y_{22}^u \epsilon^2  &  y_{23}^u \epsilon^2    \\
y_{31}^u     &  y_{32}^u      &  y_{33}^u
\end{pmatrix}, 
\M_d = \dfrac{v}{\sqrt{2}}
\begin{pmatrix}
y_{11}^d  \epsilon^5 &  y_{12}^d \epsilon^5 & y_{13}^d \epsilon^5   \\
y_{21}^d  \epsilon^3  & y_{22}^d \epsilon^3 &  y_{23}^d \epsilon^3  \\
 y_{31}^d \epsilon &  y_{32}^d \epsilon   &  y_{33}^d \epsilon
\end{pmatrix},
\M_\ell =  \dfrac{v}{\sqrt{2}}
\begin{pmatrix}
y_{11}^\ell  \epsilon^5 &  y_{12}^\ell \epsilon^5  & y_{13}^\ell \epsilon^5   \\
y_{21}^\ell  \epsilon^3  & y_{22}^\ell \epsilon^3  &  y_{23}^\ell \epsilon^3  \\
 y_{31}^\ell \epsilon   &  y_{32}^\ell \epsilon   &  y_{33}^\ell \epsilon
\end{pmatrix}.
\end{equation}

We can write similar mass matrices for the non-minimal $\mathcal{Z}_{\rm 2} \times \mathcal{Z}_{\rm 9}$ flavour symmetry \cite{Abbas:2022zfb}.

For the $\mathcal{Z}_{\rm 2} \times \mathcal{Z}_{\rm 5} $ flavour symmetry, the masses of the charged fermions turn out to be,

\begin{align}
\label{eqn5}
\{m_t, m_c, m_u\} &\simeq \{|y_{33}^u| , ~ \left |y_{22}^u- \frac {y_{23}^u y_{32}^u} {|y_{33}^u|} \right| \epsilon^2,\\&
~ \left |y_{11}^u- \frac {y_{12}^u y_{21}^u}{|y_{22}^u-y_{23}^u y_{32}^u/y_{33}^u|}- \frac{y_{13}^u |y_{31}^u y_{22}^u-y_{21}^u y_{32}^u|-y_{31}^u y_{12}^u y_{23}^u}{|y_{22}^u- y_{23}^u y_{32}^u/y_{33}^u| |y_{33}^u|} \right| \epsilon^4\}v/\sqrt{2} ,\nonumber &\\ 
\label{eqn6}
\{m_b, m_s, m_d\} & \simeq \{|y_{33}^d| \epsilon, ~ \left |y_{22}^d- \frac {y_{23}^d y_{32}^d} {|y_{33}^d|} \right| \epsilon^3,\\&
~  \left |y_{11}^d- \frac {y_{12}^d y_{21}^d}{|y_{22}^d-y_{23}^d y_{32}^d/y_{33}^d|}- \frac{y_{13}^d |y_{31}^d y_{22}^d-y_{21}^d y_{32}^d|-y_{31}^d y_{12}^d y_{23}^d}{|y_{22}^d- y_{23}^d y_{32}^d/y_{33}^d| |y_{33}^d|} \right| \epsilon^5\}v/\sqrt{2} ,\nonumber &\\
\{m_\tau, m_\mu, m_e\} & \simeq \{|y_{33}^l| \epsilon, ~ \left|y_{22}^l- \frac {y_{23}^l y_{32}^l} {|y_{33}^l|} \right| \epsilon^3,\\& ~  \left |y_{11}^l- \frac {y_{12}^l y_{21}^l}{|y_{22}^l-y_{23}^l y_{32}^l/y_{33}^l|}- \frac{y_{13}^l |y_{31}^l y_{22}^l-y_{21}^l y_{32}^l|-y_{31}^l y_{12}^l y_{23}^l}{|y_{22}^l- y_{23}^l y_{32}^l/y_{33}^l| |y_{33}^l|} \right| \epsilon^5\}v/\sqrt{2}. \nonumber &\\
\end{align}

The masses of charged fermions for the $\mathcal{Z}_{\rm 2} \times \mathcal{Z}_{\rm 9} $ flavour symmetry are,
\begin{align}
\label{eqn5}
\{m_t, m_c, m_u\} &\simeq \{|y_{33}^u| , ~ \left |y_{22}^u   \epsilon^4 - \frac {y_{23}^u y_{32}^u} {|y_{33}^u|   }  \epsilon^{12} \right|,\\&
~ \left |y_{11}^u \epsilon^8 - \frac {y_{12}^u y_{21}^u}{|y_{22}^u|} \epsilon^{10}- \frac{y_{13}^u |y_{31}^u y_{22}^u-y_{21}^u y_{32}^u|}{|y_{22}^u| |y_{33}^u|} \epsilon^8 \right| \}v/\sqrt{2} ,\nonumber &\\ 
\label{eqn6}
\{m_b, m_s, m_d\} & \simeq \{|y_{33}^d| \epsilon^3, ~ \left |y_{22}^d- \frac {y_{23}^d y_{32}^d} {|y_{33}^d|} \right| \epsilon^5,\\&
~  \left |y_{11}^d- \frac {y_{12}^d y_{21}^d}{|y_{22}^d-y_{23}^d y_{32}^d/y_{33}^d|}- \frac{y_{13}^d |y_{31}^d y_{22}^d-y_{21}^d y_{32}^d|-y_{31}^d y_{12}^d y_{23}^d}{|y_{22}^d- y_{23}^d y_{32}^d/y_{33}^d| |y_{33}^d|} \right| \epsilon^7\}v/\sqrt{2} ,\nonumber &\\
\{m_\tau, m_\mu, m_e\} & \simeq \{|y_{33}^l| \epsilon^3, ~ \left|y_{22}^l- \frac {y_{23}^l y_{32}^l} {|y_{33}^l|} \right| \epsilon^5,\\& ~  \left |y_{11}^l- \frac {y_{12}^l y_{21}^l}{|y_{22}^l-y_{23}^l y_{32}^l/y_{33}^l|}- \frac{y_{13}^l |y_{31}^l y_{22}^l-y_{21}^l y_{32}^l|-y_{31}^l y_{12}^l y_{23}^l}{|y_{22}^l- y_{23}^l y_{32}^l/y_{33}^l| |y_{33}^l|} \right| \epsilon^7\}v/\sqrt{2}. \nonumber &\\
\end{align}

The quark mixing angles for both the minimal and the non-minimal models up to the leading order are obtained as $\sin{\theta_{12}} \simeq \sin{\theta_{23}} \simeq \epsilon^2 $, and $\sin{\theta_{13}} \simeq \epsilon^4 $.

\begin{figure}[h]
	\centering
	\begin{subfigure}[]{0.45\linewidth}
	 \includegraphics[width=\linewidth]{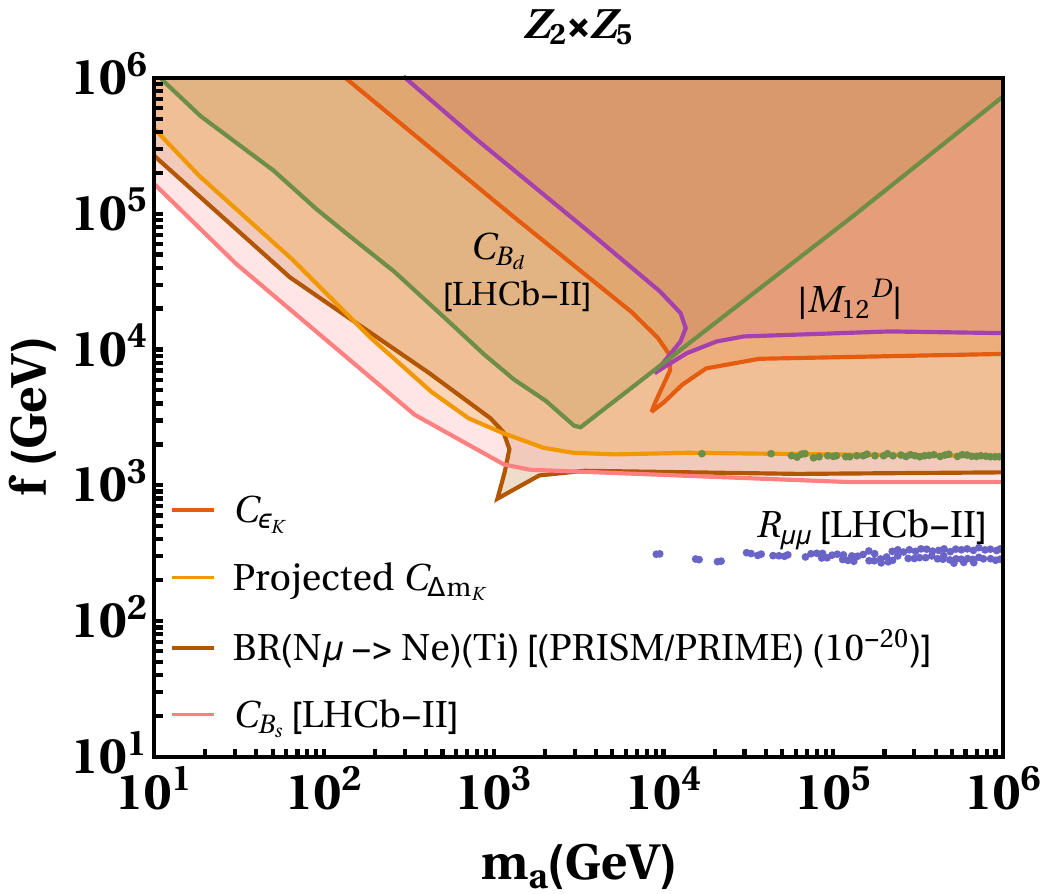}	
	   \caption{}
\end{subfigure} 
\begin{subfigure}[]{0.45\linewidth}
 \includegraphics[width=\linewidth]{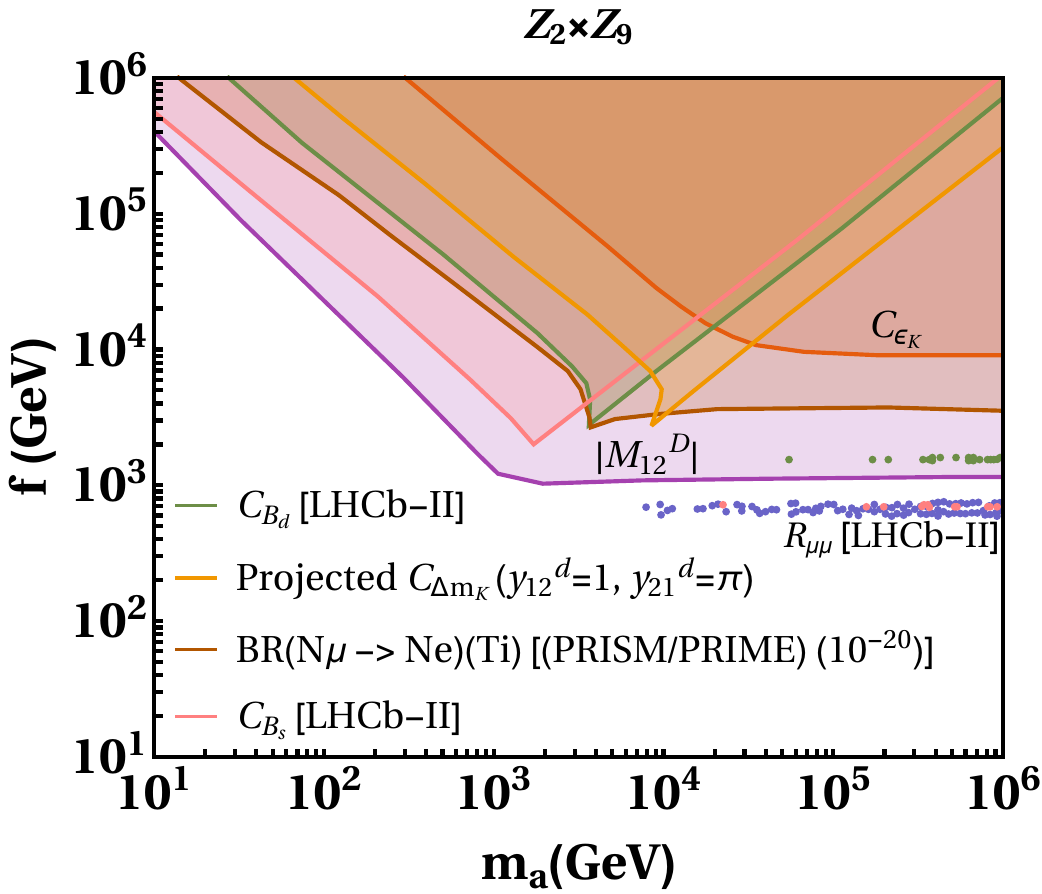}
 \caption{}
 \end{subfigure}
\caption{The bounds on the parameter space of minimal $\mathcal{Z}_2 \times \mathcal{Z}_5 $  and the non-minimal $\mathcal{Z}_2 \times \mathcal{Z}_9 $ model in the summarized form, shown in the left and the right panel, respectively. }
\label{sum_plot}
\end{figure}

Flavour bounds on these prototype symmetries are investigated in reference \cite{Abbas:2022zfb}. We parametrize the flavon field as $\chi(x)=\frac{f + s(x) +i\, a(x)}{\sqrt{2}}$, $f$ is the VEV, while $s$ and $a$ represent the scalar and pseudoscalar degrees of freedom of the flavon field.  The minimal $\mathcal{Z}_2 \times \mathcal{Z}_5$ flavour symmetry accepts $\epsilon = 0.1$, and $\epsilon = 0.23$ is used for the non-minimal  $\mathcal{Z}_2 \times \mathcal{Z}_9$ symmetry. The most stringent bounds arise from the neutral Kaon and neutral D-mixing observables.  A summary of the most important bounds on the parameter space of the minimal and non-minimal $\mathcal{Z}_2 \times \mathcal{Z}_{5,9}$ symmetries is shown in figure \ref{sum_plot}.


A detailed flavour and collider investigation of different $\mathcal{Z}_{\rm N} \times \mathcal{Z}_{\rm M}$ flavour symmetries at different hadronic colliders  is performed in reference \cite{Abbas:2024dfh}.  For instance, inclusive signatures of different $\mathcal{Z}_{\rm N} \times \mathcal{Z}_{\rm M}$ flavour symmetries can be probed at a 100 TeV collider in the process, for instance  $pp \rightarrow a \rightarrow f_i f_j$, as shown in table \ref{tab:limits_bench100a}.

\begin{table}[h!]
\setlength{\tabcolsep}{2.1pt} 
\renewcommand{\arraystretch}{1.15} 
\centering
\begin{tabular}{@{}l|rr|rr|rr|rr@{}}
\toprule
 & \multicolumn{2}{c|}{Benchmark} & \multicolumn{2}{c|}{Benchmark} & \multicolumn{2}{c|}{Benchmark} & \multicolumn{2}{c}{Benchmark}\\
 & \multicolumn{2}{c|}{$\mathcal{Z}_2 \times \mathcal{Z}_{5}$} & \multicolumn{2}{c|}{$\mathcal{Z}_2 \times \mathcal{Z}_{9}$} & \multicolumn{2}{c|}{$\mathcal{Z}_2 \times \mathcal{Z}_{11}$} & \multicolumn{2}{c}{$\mathcal{Z}_8 \times \mathcal{Z}_{22}$}\\
 $m_{a}$~[GeV] & \myalign{c}{500} & \myalign{c|}{1000} & \myalign{c}{500} & \myalign{c|}{1000} & \myalign{c}{500} & \myalign{c|}{1000} &\myalign{c}{500} & \myalign{c}{1000}\\
\midrule
$\tau \tau$~[pb]      & \fbox{$1.1 \e{-2}$} & \fbox{$1.4\e{-3}$}  & $2.2\e{-3}$ & $1.9\e{-4}$ & $4.8\e{-4}$ & $3.8\e{-5}$ & \fbox{$0.14$} & \fbox{$6.2\e{-3}$}  \\
$\mu \tau$~[pb]      & \fbox{$1.3\e{-2}$} & \fbox{$1.7\e{-3}$}  & \fbox{$6.3\e{-3}$} & \fbox{$5.5\e{-4}$} & $5\e{-4}$ & $3.9\e{-5}$ & \fbox{$0.23$} & \fbox{$1.0\e{-2}$} \\
$t  \bar{t}$~[pb]      &   &  &  & & &   & \fbox{$ 241.4$} & \fbox{$15.4$ }  \\
\bottomrule
\end{tabular}
\caption{Benchmark points for different  $\mathcal{Z}_N \times \mathcal{Z}_{M}$  flavour symmetries for inclusive flavon production channels with  flavon mass ($m_a$)  at a 100 TeV hadron collider for the flavon VEV $f = 500$ GeV.}
\label{tab:limits_bench100a}
\end{table}

\section{From flavour to dark matter}
\label{sec4}
The $\mathcal{Z}_N \times \mathcal{Z}_M$ flavour symmetry can be used to show an entirely novel scenario where a solution to the  flavour problem and  the dark matter  coexists in a single ``unique and generic"  framework \cite{Abbas:2023ion}. This can be done by writing a flavon potential which is invariant under the $\mathcal{Z}_N \times \mathcal{Z}_M$ flavour symmetry.  This requires that the flavon field $\chi$ can have maximum power $\tilde N $, which is  the least common multiple of $N$ and $M$.  Thus, we can write,
\begin{equation} \label{VN}
 V = -\lambda {\chi^{\tilde N} \over \Lambda^{\tilde N-4}} + \text{H.c.}.
\end{equation}

The mass of the axial flavon is  given by,
\begin{equation} 
\label{mphi1}
 m_\varphi^2={1\over8} |\lambda| \tilde N^2 \epsilon^{\tilde N-4} v_F^2.
\end{equation}

We note that the true vacuum and the axial flavon could be misaligned during inflation, and the range of the initial amplitude of the axial flavon lies in the range $\varphi_0=(-\pi, +\pi) v_F/\tilde N$.  This is followed by the rolling down of the axial flavon field to the true vacuum, which results in producing cold dark matter density of coherent oscillation. The equation of motion of the axial boson field amplitude in the expanding universe is, 

\begin{equation}
   \ddot{\varphi}+3H \dot{\varphi} + m_\varphi^2 \varphi \approx 0.
\end{equation}

Its  solution is $\varphi(t)=  \varphi_0  2^{1\over4} J_{1\over4}(m_\varphi t)/(m_\varphi t)^{1\over 4}$, and the  energy density  $\rho_\varphi = {1\over2}(\dot\varphi^2+m_\varphi^2 \varphi^2) $ at later time ($m_\varphi t\to \infty$) leads to  $\rho_\varphi \approx m^2_\varphi \varphi_0^2 \sqrt{2}\Gamma(5/4)^2/\pi (m_\varphi t)^{3/2}$. Using  the dark matter density, $\rho_\varphi = 0.24\, {\rm eV}^4$ at the matter-radiation equality time $t_{eq}$, that is, $m_\varphi t_{eq} \approx 2 \times 10^{27} (m_\varphi/{\rm eV})$, and equating it with the $\rho_\varphi \approx m^2_\varphi \varphi_0^2 \sqrt{2}\Gamma(5/4)^2/\pi (m_\varphi t)^{3/2}$,  we find the relation \cite{Abbas:2023ion},
\begin{equation}
\label{mphi2}
  m_\varphi =3.4\times 10^{-3} {\rm eV} \left( 10^{12} {\rm GeV} \over \varphi_0 \right)^4,
\end{equation}
which produces the right dark matter density.

Equating this with (\ref{mphi1}), we  find the relation,
\begin{equation}
\label{eq:v_F}
    v_F =2.5\times 10^7 \left( \frac{ \tilde N^{6} }{ a_0^8|\lambda| \epsilon^{\tilde N-4} } \right)^{1/10} {\rm GeV}
\end{equation}
where $\varphi_0= a_0 v_F/\tilde N$ is used. Thus, the axial flavon mass required to produce the correct dark matter density is, 
\begin{equation}
\label{flav_mass}
    m_\varphi = 0.88 \times 10^{16} \left( \epsilon^{\tilde N-4} \tilde N^4 \frac{|\lambda|}{a_0^2} \right)^{2/5} {\rm eV}.
\end{equation}

We use a model based on the $\mathcal{Z}_{8} \times \mathcal{Z}_{22}$ flavour symmetry, where  the mass of the top quark is forbidden at tree level  SM  Yukawa operator, and is generated through the dimension-5 operator.  This model is motivated by the hierarchical VEVs model \cite{Abbas:2017vws,Abbas:2020frs}, which also produces the top quark mass through the dimension-5 operator.  This model can be realized in a technicolour framework, as shown in section \ref{UV}.  The transformation of the SM and flavon fields under the  $\mathcal{Z}_{8} \times \mathcal{Z}_{22}$ flavour symmetry is defined in table \ref{tab_z6z14}. 
 \begin{table}
 \setlength{\tabcolsep}{6pt} 
\renewcommand{\arraystretch}{1.4} 
 \small
\begin{center}
\begin{tabular}{|c|c|c|c|c|c|c|c|c| c|c|c| c|c|c|}
  \hline
  Fields             &        $\mathcal{Z}_8$                    & $\mathcal{Z}_{22}$ & Fields             &        $\mathcal{Z}_8$                    & $\mathcal{Z}_{22}$   & Fields             &        $\mathcal{Z}_8$                    & $\mathcal{Z}_{22}$    & Fields             &        $\mathcal{Z}_8$                    & $\mathcal{Z}_{22}$     & Fields             &        $\mathcal{Z}_8$                    & $\mathcal{Z}_{22}$        \\
  \hline
  $u_{R}$                 &   $ \omega^2$  &$ \omega^2$        &$c_{R}$                 &   $ \omega^5$  & $ \omega^5$    &$t_{R}$                 &   $ \omega^6$  & $ \omega^6$       & $d_{R}$                 &   $ \omega^3$  &     $\omega^{3} $           & $s_{R}$                 &   $ \omega^4$  &     $\omega^4 $           \\
  $b_{R}$                 &   $ \omega^4$  &     $\omega^4 $     &   $\psi_{L,1}^q$                 &    $ \omega^2$  &    $\omega^{10} $      & $\psi_{L,2}^q$                 &  $ \omega$  &     $\omega^{9} $       &  $\psi_{L,3}^q$                 &    $\omega^{7} $  &      $\omega^{7} $ & $\psi_{L,1}^\ell$                 &   $ \omega^3$  &    $\omega^3 $          \\
     $\psi_{L,2}^\ell$                  &   $ \omega^2$  &    $\omega^2 $    &   $\psi_{L,3}^\ell$                 &   $ \omega^2$  &    $\omega^2 $     &  $e_R$                 &   $\omega^{4} $  &     $\omega^{12} $          & $\mu_R$                 &  $\omega^7 $   &     $\omega^{7} $      &  $\tau_R $                 &   $ \omega^7$  &     $\omega^{21} $              \\
            $ \nu_{e_R} $                 &     $\omega^2 $    &     $1 $         & $   \nu_{\mu_R}$                 &     $\omega^5 $    &     $\omega^{3} $          &  $  \nu_{\tau_R} $                 &     $\omega^6 $    &     $\omega^{4} $        &   $\chi$                        & $ \omega$  &       $ \omega$       & $H$              &   1        &     1                  \\          
  \hline
     \end{tabular}
\end{center}
\caption{The charges of the SM  and the flavon fields under the $\mathcal{Z}_8 \times \mathcal{Z}_{22}$  symmetry,  where $\omega$ is the 8th and  22nd root of unity. }
 \label{tab_z6z14}
\end{table}

The mass matrices of the quarks and charged leptons turn out to be,
\begin{align}
\M_u & = \dfrac{v}{\sqrt{2}}
\begin{pmatrix}
y_{11}^u  \epsilon^8 &  y_{12}^u \epsilon^{5}  & y_{13}^u \epsilon^{4}    \\
y_{21}^u \epsilon^7     & y_{22}^u \epsilon^4  &  y_{23}^u \epsilon^{3}  \\
y_{31}^u  \epsilon^{5}    &  y_{32}^u  \epsilon^2     &  y_{33}^u  \epsilon 
\end{pmatrix}, 
\M_d   = \dfrac{v}{\sqrt{2}}
\begin{pmatrix}
y_{11}^d  \epsilon^7 &  y_{12}^d \epsilon^6 & y_{13}^d \epsilon^6   \\
y_{21}^d  \epsilon^6  & y_{22}^d \epsilon^5 &  y_{23}^d \epsilon^5  \\
 y_{31}^d \epsilon^4 &  y_{32}^d \epsilon^3   &  y_{33}^d \epsilon^3
\end{pmatrix}, 
\M_\ell =  \dfrac{v}{\sqrt{2}}
\begin{pmatrix}
y_{11}^\ell  \epsilon^9 &  y_{12}^\ell \epsilon^4  & y_{13}^\ell \epsilon^4   \\
y_{21}^\ell  \epsilon^{10}  & y_{22}^\ell \epsilon^5  &  y_{23}^\ell \epsilon^3  \\
 y_{31}^\ell \epsilon^{8}   &  y_{32}^\ell \epsilon^5   &  y_{33}^\ell \epsilon^3
\end{pmatrix}.
\end{align}

The masses of charged fermions approximately read,
\begin{align}
\label{eqn5}
\{m_t, m_c, m_u\} &\simeq \{|y_{33}^u| \epsilon , ~ \left |y_{22}^u  - \frac {y_{23}^u y_{32}^u} {y_{33}^u  }   \right|  \epsilon^4 ,\\&
~ \left |y_{11}^u- \frac {y_{12}^u y_{21}^u}{y_{22}^u-y_{23}^u y_{32}^u/y_{33}^u}- \frac{y_{13}^u (y_{31}^u y_{22}^u-y_{21}^u y_{32}^u)-y_{31}^u y_{12}^u y_{23}^u}{(y_{22}^u- y_{23}^u y_{32}^u/y_{33}^u) y_{33}^u} \right| \epsilon^8\}v/\sqrt{2}  ,\nonumber \\ 
\{m_b, m_s, m_d\} & \simeq \{|y_{33}^d| \epsilon^3, ~ \left |y_{22}^d- \frac {y_{23}^d y_{32}^d} {y_{33}^d} \right| \epsilon^5,\\ \nonumber 
&  \left |y_{11}^d- \frac {y_{12}^d y_{21}^d}{y_{22}^d-y_{23}^d y_{32}^d/y_{33}^d}- \frac{y_{13}^d (y_{31}^d y_{22}^d-y_{21}^d y_{32}^d)-y_{31}^d y_{12}^d y_{23}^d}{(y_{22}^d- y_{23}^d y_{32}^d/y_{33}^d) y_{33}^d} \right| \epsilon^7\}v/\sqrt{2} ,\\ \nonumber 
\{m_\tau, m_\mu, m_e\} & \simeq \{|y_{33}^l| \epsilon^3, ~ \left|y_{22}^l- \frac {y_{23}^l y_{32}^l} {y_{33}^l} \right| \epsilon^5,\\& ~  \left |y_{11}^l- \frac {y_{12}^l y_{21}^l}{y_{22}^l-y_{23}^l y_{32}^l/y_{33}^l}- \frac{y_{13}^l \left( y_{31}^l y_{22}^l-y_{21}^l y_{32}^l \right) -y_{31}^l y_{12}^l y_{23}^l}{\left(  y_{22}^l- y_{23}^l y_{32}^l/y_{33}^l \right) y_{33}^l} \right| \epsilon^9\}v/\sqrt{2}.
\end{align}
The mixing angles of quarks are given by,
\begin{eqnarray}
\sin \theta_{12}  \simeq |V_{us}| &\simeq& \left|{y_{12}^d \over y_{22}^d}  -{y_{12}^u \over y_{22}^u}  \right| \epsilon, ~
\sin \theta_{23}  \simeq |V_{cb}| \simeq  \left|{y_{23}^d \over y_{33}^d}   -{y_{23}^u \over y_{33}^u}   \right| \epsilon^2,~
\sin \theta_{13}  \simeq |V_{ub}| \simeq  \left|{y_{13}^d \over y_{33}^d}    -{y_{12}^u y_{23}^d \over y_{22}^u y_{33}^d}      
- {y_{13}^u \over y_{33}^u}   \right|   \epsilon^3. \qquad
\end{eqnarray}
The Dirac CP phase is predicted to be  $\delta = 1.196$ \cite{Abbas:2024dfh}.  For $\tilde N=88$ corresponding to  the  $\mathcal{Z}_{8} \times \mathcal{Z}_{22}$ flavour symmetry, we obtain,
\begin{align}
v_F  \approx 1.0 \times 10^{14}\, \rm{GeV}, ~~\text{and}~~
m_a \approx   1.9 \times 10^{-3} \,  \rm{eV},
\end{align}
where $\epsilon=0.225$ with $|\lambda|=1$ and $a_0=1$ is considered.

We can choose another symmetry without deviating from the flavour structure produced by the  $\mathcal{Z}_{8} \times \mathcal{Z}_{22}$ flavour symmetry.  For instance, a smaller symmetry $\mathcal{Z}_{4} \times \mathcal{Z}_{17}$ with $\tilde N=68$ provides $v_F  \approx 4.4 \times 10^{12}\, \rm{GeV}$ and $m_a \approx  196 \,  \rm{eV}$.  Thus, our framework is rather generic.  For the stability of the flavonic dark matter, it is necessary that it should not decay into a pair of electrons and positrons.  This requirement places a stringent bound on the value of $\tilde N$ given by,
\begin{equation} \label{Nmin}
\tilde N > 53, ~~\mbox{and} ~~ v_F > 4 \times 10^{11} \mbox{GeV}.
\end{equation}

\begin{figure}
	\centering
	 \includegraphics[width= 0.6\linewidth]{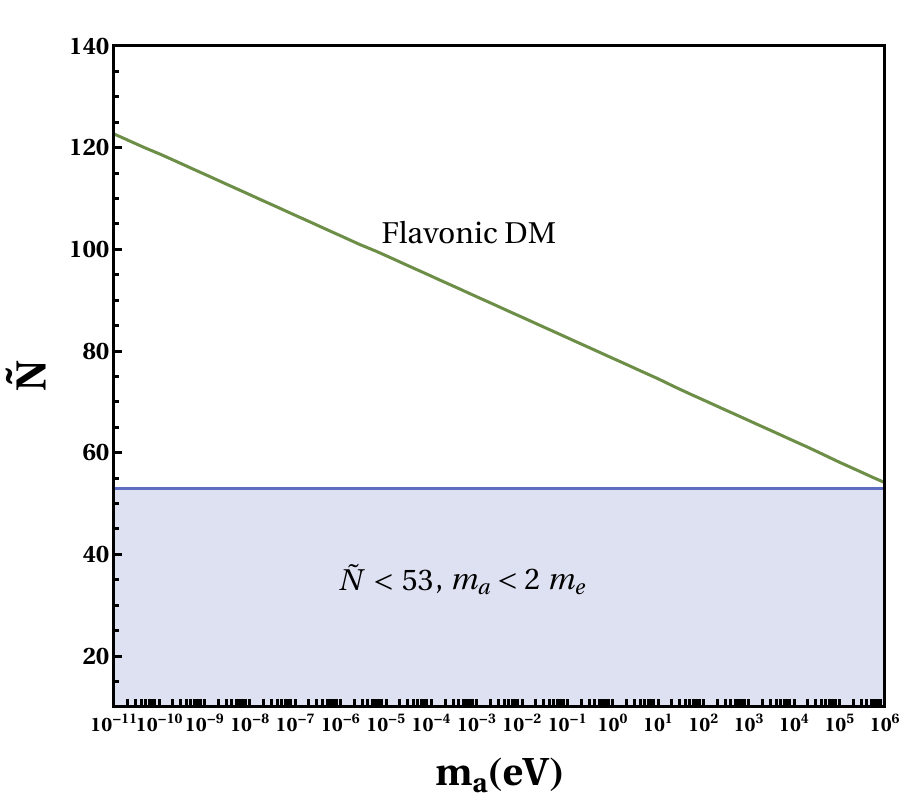}	
\caption{ The allowed mass range of flavonic dark matter for a range of $\tilde N$.  The light-blue region is ruled out by the requirement $m_a < 2 m_e$.  }
\label{n_ma}
\end{figure}

We show a range of $\tilde N$ in figure \ref{n_ma} where the light-blue region is excluded by the demand $m_a < 2 m_e$.

The  flavon scale $v_F$ is bounded by  the FCNC process $K^+ \to \pi^+ a $ \cite{Bjorkeroth:2018dzu}:
\begin{eqnarray}
    v_F \gtrsim 7 \times 10^{11} V^d_{21} {\rm GeV}, 
\end{eqnarray}
where we have $V^d_{21} \approx \epsilon$. 

The coupling of the flavon $a$ with a pair of fermion and anti-fermions has a generic structure $ v m_f/ \sqrt{2} v_F$ where $m_f$ is the mass of the SM fermion.  Since $v_F \gtrsim 7 \times 10^{11} V^d_{21} {\rm GeV}$, these couplings are extremely suppressed due to the scale $v_F$ appearing in the denominator.  This causes an extreme suppression of the decay of the flavon $a$ to the SM fermions by a factor of $1/v_F^2$.  For instance,   the decay rate of the process $ a \to \nu \nu$, for $\tilde N=88$,  turns out to be \cite{djouadi},
  \begin{eqnarray}
 \Gamma (a \to \nu \nu) =  \frac{1}{8\pi} g_{a\nu\nu}^2 m_a \beta_\nu \approx 9.95 \times 10^{-60} \rm GeV,
  \end{eqnarray}
where $\beta_f = (1-4 m_\nu^2/m_a^2)^{1/2}$ and $g_{a\nu\nu} = 20 \frac{v}{ v_F \sqrt{2}} \epsilon^{20}$.  
   
The decay  $a \rightarrow \gamma \gamma$ occurs through a top-quark triangle-loop, and its decay-width is given by, \cite{djouadi,Spira:2016ztx}, 
\begin{eqnarray}
\Gamma (a \rightarrow \gamma \gamma)=\frac{\alpha ^2 m_{a}^3}{64 \pi ^3 v_F^2}\left| \sum_f N_{c f}Q_f^2 F_{1/2}(\tau ) \right|^2= 1.69 \times 10^{-68} \rm GeV,
\label{diphoton-width}
\end{eqnarray}
where $\tau=\frac{4 m_{f}^2}{m_{\varphi}^2}$, $F_{1/2} = \tau f(\tau )$, $f(\tau )= \left[\sin ^{-1}(1/\sqrt{\tau })\right]^2 $, and $Q_f$ and $N_{c f}$ are the electric charge  and  the color factor corresponding to the fermion $f$. Thus, we see that the decay widths of these processes  correspond to the lifetime of flavonic dark matter much larger than the age of the universe.

The  coupling of the flavonic dark matter to photons is the striking channel to observe it,
\begin{align}
    \mathcal{L}^{a \gamma \gamma}_{\rm eff} = \frac{1}{4} g_{a\gamma\gamma} \varphi  F^{\mu\nu} \Tilde{F}_{\mu \nu}.
\end{align}

\begin{figure}[h]
    \centering
\includegraphics[width= 0.7\linewidth]{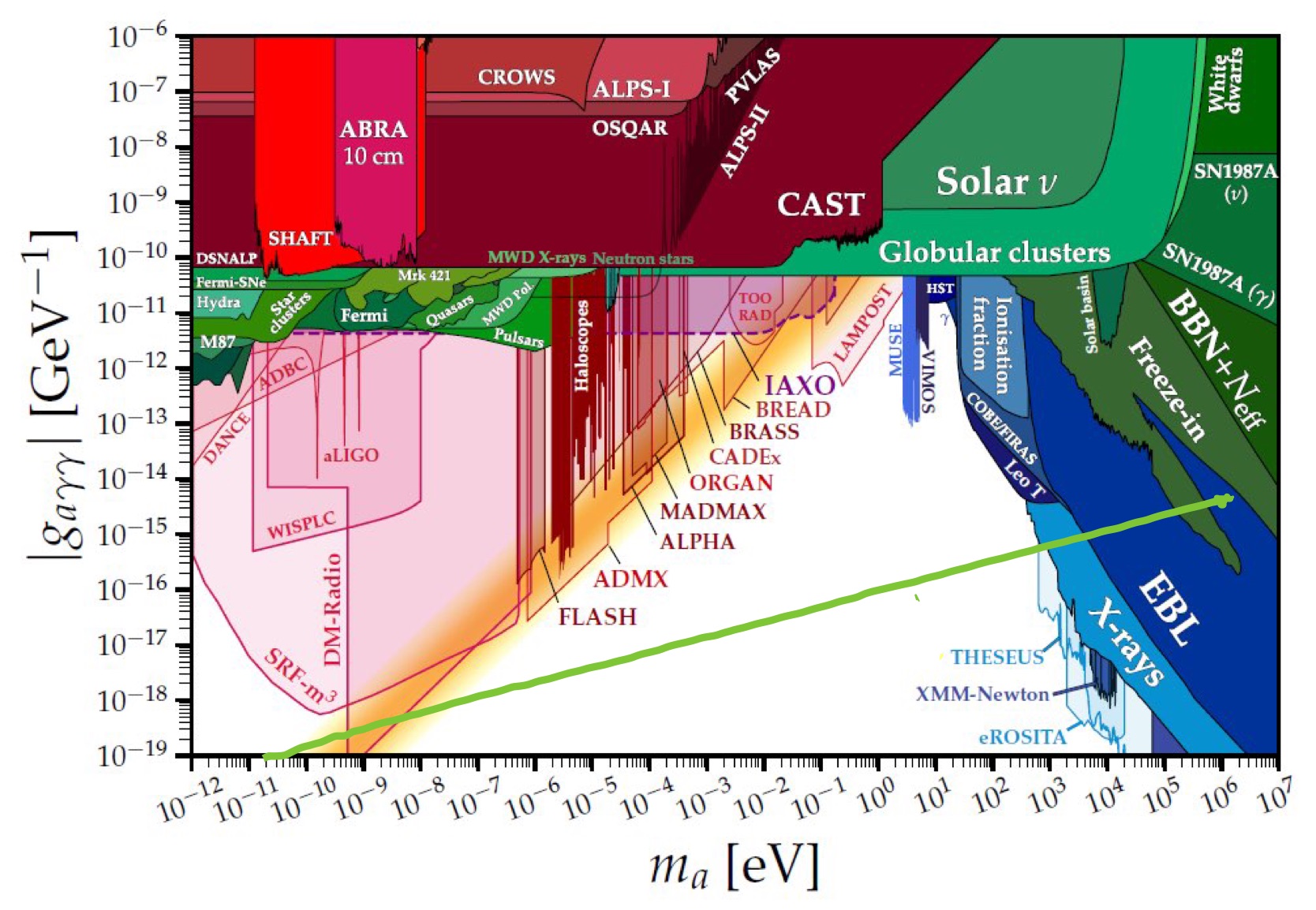} 
  \caption{ The prediction of flavonic dark matter (thick green line) and axion-like particle  searches \cite{Antel:2023hkf}. }
    \label{fig:fips}
\end{figure}
The predictions of the  coupling of the flavonic dark matter to photons are shown in figure \ref{fig:fips} by the thick green line, which is overlaid in Fig.~15 of \cite{Antel:2023hkf}.  The flavonic dark matter mass corresponding to $\tilde N < 67$ and $v_F < 4 \times 10^{12}$ GeV, which is greater than about 1 KeV, is ruled out.  This is in agreement with the latest measurement by INTEGRAL/SPI \cite{INT}. The forthcoming experiment THESEUS  is expected to further examine the above KeV mass range  \cite{Thorpe-Morgan:2020rwc}.  We notice that a small range  of the flavonic dark matter mass overlaps with that of the GUT-scale QCD axion at around $10^{-9}$ eV.  This can be probed in the future \cite{DMRadio:2022jfv}.  We also notice that there is a very small overlap between the mass of the flavonic dark matter and QCD-axion (denoted by orange band) between the approximate mass range $10^{-10}-10^{-8}$ eV as shown in figure \ref{fig:fips}.

\section{The standard HVM}
\label{sec:HVM}
The hierarchical VEVs models (HVM) present a radical approach to the flavour problem by introducing the six gauge singlet scalar fields $\chi_r$ whose VEVs account for the fermionic mass spectrum and flavour mixing \cite{Abbas:2017vws,Abbas:2020frs,Abbas:2023dpf,Abbas:2023bmm}. The central idea of this approach is that the gauge singlet scalar fields $\chi_r$ are not fundamental scalars, and are ``multi-fermions bound states"  of a dark-technicolour dynamics, which will be discussed in the next section.

The composite scalar fields  $\chi _i $  under the SM symmetry $\mathcal{G}_{\rm SM} \equiv SU(3)_c \otimes SU(2)_L \otimes U(1)_Y$ transform as,
\begin{eqnarray}
\chi_r :(1,1,0),
 \end{eqnarray} 
where $r=1-6$.

The HVM redefines the flavour problem in terms of the VEVs hierarchy instead of the hierarchical Yukawa couplings and then solves it \cite{Abbas:2017vws,Abbas:2020frs,Abbas:2023dpf,Abbas:2023bmm,Abbas:2024jut}.  Apparently, this does not seem a solution to the flavour problem, since the hierarchy of the Yukawa couplings is replaced by the hierarchy of the VEVs, which are still free parameters of the theory.   However, we hope that nature plays a subtle role in the case of flavour puzzle.  The loftiness of nature may be revealed in an underlying theory where all the different hierarchical VEVs are multi-fermions chiral condensates of a dark-technicolour dynamics, and can be  parameterized in terms of a single parameter of the dark-technicolour interactions.    Such a scenario will be discussed in section \ref{UV}  of this review.

In this work, we discussed what we call the ``standard" HVM (SHVM) scenario, which is capable of predicting precise leptonic mixing angles in terms of the Cabibbo angle and the mass of the charm quark \cite{Abbas:2023dpf}.  The SHVM is obtained by imposing a generic   $\mathcal{Z}_{\rm N} \times \mathcal{Z}_{\rm M} \times \mathcal{Z}_{\rm P}$ flavour symmetry on the SM, which naturally arises in the dark-technicolour paradigm through the breaking of three axial $U(1)_{A}$ symmetries  \cite{Abbas:2020frs,Abbas:2023bmm}.

The SHVM shows an emergence of a multiple ALPs scenario \cite{Abbas:2023dpf,Abbas:2024jut}.  In general,  ALPs may be a cold dark matter candidate, and may even provide a solution to the strong CP problem if they are axions.  The  multiple ALPs scenario can also occur in the framework of  string theory, where it is called an axiverse, that is, an existence of a large number of ALPs \cite{Svrcek:2006yi,Cicoli:2012sz,Broeckel:2021dpz,Demirtas:2021gsq,Gendler:2023kjt}. Thus, the SHVM may be a low energy limit of a string-type dynamics.   

Moreover, multiple ALPs scenario can also appear in a field theory axiverse ($\pi$ axiverse ), which is an alternative to the string theory axiverse.  The field theory axiverse arises from a  dark QCD where  $N_f$ flavours of dark-quarks produce $N_f^2-1$ axionlike states \cite{Bai:2013xga,Tsai:2020vpi,Alexander:2020wpm}.  In the UV completion of the SHVM, there are three  strong QCD-like gauge symmetries, which are $SU(N)_{TC}$, $SU(N)_{DTC}$, and $SU(N)_{F}$, where TC is for technicolour, DTC stands for a dark-technicolour, and F shows the strong dynamics of vector-like fermions.  Thus, the SHVM embedded in the dark-technicolour paradigm is a practical example of a field theory axiverse. Multiple ALPs scenario is recently investigated in reference \cite{Chadha-Day:2023wub}.

We notice that due to presence of the hierarchical VEVs, the scalar mass spectrum of the SHVM is also hierarchical.  This fact may be used to address several anomalies in table \ref{tab:anomalies}.  Moreover, since  the SHVM is embedded in the dark-technicolour framework, we can have $\rho (\chi_r^2 + \chi_r^{\dagger 2})$ type soft-symmetry breaking terms in scalar potential of the SHVM.  These terms can be dynamically generated from the strong dark-technicolour  dynamics.  Thus,   the ALPs of the SHVM may become heavier through the $\rho (\chi_r^2 + \chi_r^{\dagger 2})$ type soft-symmetry breaking terms, and can be looked for at the LHC.  Moreover, in this scenario, they may account for pseudoscalar anomalies given in table \ref{tab:anomalies}.

 The masses of the charged-fermions originate from  the  dimension-5 operators as given below,
\bea
\label{mass2}
{\mathcal{L}} &=& \dfrac{1}{\Lambda }\Bigl[  y_{ij}^u  \bar{\psi}_{L_i}^{q}  \tilde{\varphi} \psi_{R_i}^{u}   \chi_r +     
   y_{ij}^d  \bar{\psi}_{L_i}^{q}   \varphi \psi_{R_i}^{d}  \chi_r   +   y_{ij}^\ell  \bar{\psi}_{L_i}^{\ell}   \varphi \psi_{R_i}^{\ell}  \chi_r \Bigr]  
+  {\rm H.c.}.
\eea

For producing the required pattern of the charged fermion masses, the left-handed fermionic doublet must have the following charges under the generic   $\mathcal{Z}_{\rm N} \times \mathcal{Z}_{\rm M} \times \mathcal{Z}_{\rm P}$ flavour symmetry,
\begin{align}
\psi_{L_1}^{q} &: (+, 1, \omega^{13}),~ \psi_{L_2}^{q}: (+, 1, \omega^{2 n +1}),~  \psi_{L_3}^{q}: (+, 1, \omega^{2 n+ 2}), \\ \nonumber
\psi_{L_1}^{\ell} &: (+, 1, \omega^{12}),~ \psi_{L_2}^{\ell}: (+, 1, \omega^{10 -n}),~  \psi_{L_3}^{\ell}: (+, 1, \omega^{6}),
\end{align}
where $n=0,1,2 \cdots$, and $\rm N= 2$, $\rm M\geq 4$,  and $\rm P \geq 14$ are required.  We observe that for recovering the charged fermionic mass pattern, we must fix $\rm N= 2$.

The neutrino masses are obtained  by adding   three right-handed  neutrinos $\nu_{eR}$, $\nu_{\mu R}$, $\nu_{\tau R}$  to the SM, and by writing the dimension-6 operators as,
\begin{eqnarray}
\label{mass_N}
-{\mathcal{L}}_{\rm Yukawa}^{\nu} &=&      y_{ij}^\nu \bar{ \psi}_{L_i}^\ell   \tilde{\varphi}  \nu_{f_R} \left[  \dfrac{ \chi_r \chi_7 (\text{or}~ \chi_r  \chi_7^\dagger)}{\Lambda^2} \right] +  {\rm H.c.}. 
\end{eqnarray}

The demand that the neutrino masses arise from equation \ref{mass_N} imposes a severe constraint on the symmetry  $\mathcal{Z}_{\rm P}$, allowing $\rm P = 14$ only\footnote{We do not find any specific reason for this within the effective model.   We hope to investigate this point  further in a UV completion based on the dark-technicolour.}.  The rest of the fields transform under the  $\mathcal{Z}_2 \times \mathcal{Z}_4 \times \mathcal{Z}_{14} $ flavour symmetry, as given  in table  \ref{tab1}. We explain the  fermionic mass hierarchy in terms of  the   VEVs pattern    $ \langle \chi _4 \rangle > \langle \chi _1 \rangle $, $ \langle \chi _2 \rangle >> \langle \chi _5 \rangle $, $ \langle \chi _3 \rangle >> \langle \chi _6 \rangle $, $ \langle \chi _{3} \rangle >> \langle \chi _{2} \rangle >> \langle \chi _{1} \rangle $, and  $ \langle \chi _6 \rangle >> \langle \chi _5 \rangle >> \langle \chi _4 \rangle $.  

\begin{table}[H]
\begin{center}
\resizebox{\textwidth}{!}{
\begin{tabular}{|c|c|c|c||c|c|c|c||c|c|c|c||c|c|c|c|c|}
  \hline
  Fields                               &   $\mathcal{Z}_2$  &  $\mathcal{Z}_4$   &  $\mathcal{Z}_{14}$   & Fields   &  $\mathcal{Z}_2$   &  $\mathcal{Z}_4$ &  $\mathcal{Z}_{14}$ & Fields   & $\mathcal{Z}_2$  & $\mathcal{Z}_4$  &  $\mathcal{Z}_{14}$  & Fields  &  $\mathcal{Z}_2$   &  $\mathcal{Z}_4$  &  $\mathcal{Z}_{14}$ \\
  \hline
 $u_{R}$                        &     -   &     $ 1$          &    $\omega^{11}$ & $d_{R} $, $ s_{R}$, $b_{R}$    &     +    & $ 1$        &    $\omega^{12}$  & $ \psi_{L_3}^{q} $       &    +     &  $ 1$    &   $\omega^2$   &  $\tau_R$      &   +  &  $\omega^3$      &     $\omega$             \\
  $c_{R}$                       &     +   &    $ 1$          &    $\omega^6$   &  $\chi _4$                         &      +  &  $ 1$      &  $\omega^{13}$  &  $ \psi_{L_1}^\ell $                          &     +   & $\omega^3$     &  $\omega^{12}$                          &   $\nu_{e_R}$   &    +   &   $\omega$     &      $\omega^{8}$        \\
   $t_{R}$                        &     +   &    $1$         &    $\omega^{4}$   & $\chi _5$                         &      +  &  $1$    &  $\omega^{11}$  &  $ \psi_{L_2}^{\ell} $     &      +  & $\omega^3$      &  $\omega^{10}$     & $\nu_{\mu_R}$                   &     -  &   $\omega$    &   $\omega^3$                    \\
  $\chi _1$                        &      -  &   $ 1 $      &    $\omega^2$    &   $\chi _6$                          &      +  &  $ 1$      &   $ \omega^{10}$    &   $ \psi_{L_3}^{ \ell} $       &    +     &  $\omega^3$    &   $\omega^6$                                 & $\nu_{\tau_R}$                    &     -   &  $\omega$     &   $\omega^3$          \\
  $\chi _2$                   & +     &       $ 1$      &  $\omega^5$   & $ \psi_{L_1}^q $                          &      +  &  $1$      &  $\omega^{13}$  & $e_R$    &      -   &   $\omega^3$       &    $\omega^{10}$      &  $\chi_7 $                          &      -   &  $\omega^2$   &     $\omega^8$                                               \\
   $\chi _3$                  &    +   &       $ 1$    & $ \omega^2$        & $ \psi_{L_2}^{q} $     &      +  & $1 $      &  $\omega$  &   $ \mu_R$     &   +  & $\omega^3$       &     $\omega^{13}$      &  $ \varphi $                           &      +  &1     &   1                                            \\
  \hline
     \end{tabular}}
\end{center}
\caption{The charges of left- and right-handed fermions  and  scalar fields under $\mathcal{Z}_2$, $\mathcal{Z}_4$,  and $\mathcal{Z}_{14}$ symmetries for the normal mass ordering. The $\omega$ is the 4th and 14th root of unity corresponding to the symmetries $\mathcal{Z}_4$  and $\mathcal{Z}_{14}$,  respectively.}
 \label{tab1}
\end{table}

The mass matrices of up, down-type quarks and leptons are,
\begin{align}
\label{mUD}
\M_\U & =   \dfrac{ v }{\sqrt{2}} 
\begin{pmatrix}
y_{11}^u  \epsilon_1 &  0  & y_{13}^u  \epsilon_2    \\
0    & y_{22}^u \epsilon_2  & y_{23}^u  \epsilon_5   \\
0   &  y_{32}^u  \epsilon_6    &  y_{33}^u  \epsilon_3 
\end{pmatrix},  
\M_\D = \dfrac{ v }{\sqrt{2}} 
 \begin{pmatrix}
  y_{11}^d \epsilon_4 &    y_{12}^d \epsilon_4 &  y_{13}^d \epsilon_4 \\
  y_{21}^d \epsilon_5 &     y_{22}^d \epsilon_5 &   y_{23}^d \epsilon_5\\
    y_{31}^d \epsilon_6 &     y_{32}^d \epsilon_6  &   y_{33}^d \epsilon_6\\
\end{pmatrix},  
\M_\ell  =\dfrac{ v }{\sqrt{2}} 
  \begin{pmatrix}
  y_{11}^\ell \epsilon_1 &    y_{12}^\ell \epsilon_4  &   y_{13}^\ell \epsilon_5 \\
 0 &    y_{22}^\ell \epsilon_5 &   y_{23}^\ell \epsilon_2\\
   0  &    0  &   y_{33}^\ell \epsilon_2 \\
\end{pmatrix},
\end{align} 
where $\epsilon_r = \dfrac{\langle \chi_r \rangle }{\Lambda}$ and  $\epsilon_r<1$.  

The masses of charged fermions read,
\begin{eqnarray}
\label{mass_charged}
m_t  &\approx& \ \left|y^u_{33} \right| \epsilon_3 v/\sqrt{2}, ~
m_c  \approx \   |y^u_{22} \epsilon_2  |  v /\sqrt{2} ,~
m_u  \approx  |y_{11}^u  |\,  \epsilon_1 v /\sqrt{2},
m_b  \approx \ |y^d_{33}| \epsilon_6 v/\sqrt{2}, 
m_s  \approx \   |y^d_{22}  | \epsilon_5 v /\sqrt{2} ,\nonumber \\
m_d  &\approx&  |y_{11}^d    |\,  \epsilon_4 v /\sqrt{2},
m_\tau  \approx \ |y^\ell_{33}| \epsilon_2 v/\sqrt{2}, ~
m_\mu  \approx \   |y^\ell_{22} | \epsilon_5 v /\sqrt{2} ,~
m_e  =  |y_{11}^\ell   |\,  \epsilon_1 v /\sqrt{2}.
\end {eqnarray}

We can write the quark mixing angles as,
\begin{eqnarray}
\sin \theta_{12}  & \simeq&  \left|{y_{12}^d \epsilon_4 \over y_{22}^d \epsilon_5}   \right|= { \epsilon_4 \over \epsilon_5}, ~
\sin \theta_{23}  \simeq  \left| {y_{23}^d \epsilon_5  \over y_{33}^d \epsilon_6 }  - {y_{23}^u \epsilon_5  \over y_{33}^u \epsilon_3 }   \right|,~
\sin \theta_{13}  \simeq   \left|{y_{13}^d  \epsilon_4  \over y_{33}^d  \epsilon_6 } -{y_{13}^u  \epsilon_2  \over y_{33}^u  \epsilon_3 } \right|.
\end{eqnarray} 

Assuming all $|y_{ij}^{d}| =1 $,  the quark mixing angles can be written as,
\begin{eqnarray}
\sin \theta_{12}  & \simeq&   { \epsilon_{4} \over \epsilon_{5}}, ~ \\ \nonumber 
\sin \theta_{23} & \simeq &  { \epsilon_{5} \over \epsilon_{6}}, \\ \nonumber 
\sin \theta_{13}  &\simeq &  { \epsilon_{4} \over \epsilon_{6}} - \left|\frac{ y_{13}^u}{ y_{33}^u} \right| { \epsilon_{2} \over \epsilon_{3}}.
\end{eqnarray} 

This leads  to  the prediction of the quark mixing angle $ \theta_{13} $ as,
\begin{eqnarray}
\sin \theta_{13}  \simeq   \sin \theta_{12}  \sin \theta_{23}  -  { \epsilon_{2} \over \epsilon_{3}},
\end{eqnarray} 
assuming $|y_{ij}^{u}| =1 $.  We can predict quark mixing angle $\sin \theta_{13}$ for an allowed value $\epsilon_3= 0.55$.  Moreover, we predict the Dirac CP phase  to be $\delta = 1.196$. 

In general, the $\epsilon_r$  parameters are \cite{Abbas:2023dpf},
\begin{equation}
\label{epsi}
\epsilon_1 = 3.16 \times 10^{-6},~ \epsilon_2 = 0.0031,~ \epsilon_3 = 0.87,~\epsilon_4 = 0.000061,~\epsilon_5 = 0.000270,~\epsilon_6 = 0.0054,~\epsilon_7 = 7.18 \times 10^{-10}.  
\end{equation}

We have only the Dirac neutrinos in the SHVM, the mass matrix for neutrinos becomes,
\begin{equation}
\label{NM}
\M_{\N} = \dfrac{v}{\sqrt{2}}  
\begin{pmatrix}
y_{11}^\nu   \epsilon_1 \epsilon_7   &  y_{12}^\nu   \epsilon_4 \epsilon_7  & y_{13}^\nu  \epsilon_4  \epsilon_7 \\
0   & y_{22}^\nu  \epsilon_4  \epsilon_7 &  y_{23}^\nu  \epsilon_4  \epsilon_7 \\
0   &   y_{32}^\nu  \epsilon_5  \epsilon_7   &  y_{33}^\nu  \epsilon_5  \epsilon_7
\end{pmatrix}.
\end{equation}

The neutrino masses turn out to be,
\begin{eqnarray}
\label{mass_neutrino}
m_3  &\approx&  |y^\nu_{33}|  \epsilon_5 \epsilon_7 v/\sqrt{2}, 
m_2  \approx     |y^\nu_{22} - \dfrac{y_{23}^\nu  y_{32}^\nu}{y_{33}^\nu} |  \epsilon_4 \epsilon_7 v /\sqrt{2},
m_1  \approx  |y_{11}^\nu  |\,  \epsilon_1 \epsilon_7 v /\sqrt{2}, \quad \quad
\end {eqnarray}
and leptonic mixing angles are,
\begin{eqnarray}
\sin \theta_{12}^\ell  \approx  1 - 2 \sin \theta_{12}, 
\sin \theta_{23}^\ell  \approx  1 -  \sin \theta_{12}, 
\sin \theta_{13}^\ell   \approx \sin \theta_{12} - \frac{m_s}{m_c}.
\end{eqnarray}

Phenomenological investigation of the SHVM is recently performed in reference \cite{Abbas:2024jut}, where flavour bounds are derived for the symmetry conserving scalar potential, and collider physics is investigated for the softly-broken scalar potential.  Moreover, it is shown that  pseudoscalar $a_3$, in the scenario where the masses of pseudoscalars are much lighter than that of scalars,  pseudoscalar $a_3$ can account for 95.4 GeV excess given in table  \ref{tab:anomalies}.  However, this is a very specific case of the SHVM, and in more general scenario, the SHVM may accommodate several anomalous masses, given in table \ref{tab:anomalies}\footnote{Investigation is under progress.}. 

Furthermore, the  axial degree of freedom of the field $\chi_7$,  denoted by $a_7$,  may be a new class of   dark matter, since it only couples to neutrino among the fermions. Its mass can be generated through the soft symmetry breaking potential given by \cite{Abbas:2023dpf},
\begin{align}
V_{\rm soft}^{a_7} =  \rho_7 (\chi_7^2 + \chi_7^{*2}).
\end{align}
The mass of the  $a_7$ can be written as,
\begin{align}
m_{a_7} = \sqrt{\rho_7},
\end{align}
which turns out to be a free parameter of the SHVM.  The decay rate of  $a_7$ to apair of neutrinos is  \cite{djouadi},
\begin{eqnarray}
 \Gamma (a_7 \to \nu \nu) =  \frac{1}{8\pi} g_{a_7 \nu\nu}^2 m_{a_7} \beta_\nu,
  \end{eqnarray}
  where $\beta_f = (1-4 m_\nu^2/m_{a_7}^2)^{1/2}$ and $g_{a_7\nu\nu} =  \frac{v}{ v_7 \sqrt{2}} \epsilon_{5} \epsilon_7 =    \frac{v}{ 2 \Lambda} \epsilon_{5}$. For  $m_{a_7} = 10^{-8}\rm GeV$, the $\Gamma (a_7 \to \nu \nu)$ is of the order $8.8 \times 10^{-45} $ GeV for the scale $\Lambda = 10^{16}$ GeV.  However, if we have $m_{a_7}< 2 m_\nu$, the scale $\Lambda$ can be as low as the eletroweak scale, and $a_7$ may still be a possible dark-matter particle if the mass is greater than  $10^{-21}$ eV \cite{Cirelli:2024ssz}.

\section{The dark-technicolour paradigm: beyond flavour and dark matter}
\label{UV}
In this section, we discuss a common  UV origin of the  models based on the $\mathcal{Z}_{\rm N} \times \mathcal{Z}_{\rm M}$ flavour symmetry, and the VEVs hierarchy that is based on a generic   $\mathcal{Z}_{\rm N} \times \mathcal{Z}_{\rm M} \times \mathcal{Z}_{\rm P}$ flavour symmetry.  For this purpose, we use the technicolour scenario proposed in reference \cite{Abbas:2020frs}.  In this scenario, we have the   $\mathcal{G} \equiv SU(\rm N_{\rm TC}) \times SU(\rm N_{\rm DTC}) \times SU(\rm{N}_{\rm F})$ symmetry, where TC stands for technicolour,  DTC for dark-technicolour, and F represents strong dynamics of vector-like fermions.

In the TC dynamics, there are $\rm K_{\rm TC}$ flavours behaving  under the $SU(3)_c \times SU(2)_L \times U(1)_Y \times \mathcal{G}$ as \cite{Abbas:2020frs},
\begin{eqnarray}
T_{q}^i  &\equiv&   \begin{pmatrix}
T  \\
B
\end{pmatrix}_L:(1,2,0,\rm{N}_{\rm TC},1,1), ~
T_{R}^i : (1,1,1,\text{N}_{\rm{TC}},1,1), B_{R}^i : (1,1,-1,\rm{N}_{\rm TC},1,1), 
\end{eqnarray}
where $i=1,2,3 \cdots $,  and the electric charge of $T$ is  $+\frac{1}{2}$  and $-\frac{1}{2}$ for $B$, and  the first three quantum numbers refer to the SM.

The dark TC symmetry $SU(\rm N_{\rm DTC}) $ has $\rm K_{\rm DTC}$ flavours  transforming under  $SU(3)_c \times SU(2)_L \times U(1)_Y \times \mathcal{G}$ as \cite{Abbas:2020frs},
\begin{eqnarray}
 \mathcal{D}_{ q}^i &\equiv& \mathcal{C}_{L,R}^i  : (1,1, 1,1,\text{N}_{\text{DTC}},1),~\mathcal{S}_{L,R}^i  : (1,1,-1,1,\text{N}_{\text{DTC}},1), 
\end{eqnarray}
where  $i=1,2,3 \cdots $,  and  electric charges are $+\frac{1}{2}$ for $\mathcal C$ and $-\frac{1}{2}$ for $\mathcal S$.  

In a similar manner, there are   $\rm K_{\rm F}$  fermionic flavours corresponding to the $SU(N_{\text{F}})$ symmetry transforming under   $SU(3)_c \times SU(2)_L \times U(1)_Y \times \mathcal{G}$ as \cite{Abbas:2020frs},
\begin{eqnarray}
F_{L,R} &\equiv &U_{L,R}^i \equiv  (3,1,\dfrac{4}{3},1,1,\text{N}_\text{F}),
D_{L,R}^{i} \equiv   (3,1,-\dfrac{2}{3},1,1,\text{N}_\text{F}),  \\ \nonumber 
N_{L,R}^i &\equiv&   (1,1,0,1,1,\text{N}_\text{F}), 
E_{L,R}^{i} \equiv   (1,1,-2,1,1,\text{N}_\text{F}),
\end{eqnarray}
where  $i=1,2,3 \cdots $.  
 
\subsection{UV completion of the SHVM}
Now we assume the existence of an extended technicolour  dynamics (ETC) mediating the TC fermions,  the left-handed SM fermions, and the $F_R$ fermions, and an extended-DTC (EDTC) symmetry containing the  DTC fermions,  the right-handed SM singlet fermions, and the $F_L$ fermions.  We notice that the $SU(\rm N)_{\rm F}$ symmetry acts like a connecting bridge between the TC and DTC dynamics and results in a suppression of the mixing between the TC and the DTC dynamics by the factor $1 / \Lambda$.  This may explain the SM-like behaviour of the  discovered Higgs.

There are three axial  $U(1)_{\rm A}$  symmetries, namely,   $U(1)_{\rm A}^{\rm TC }$,   $U(1)_{\rm A}^{\rm DTC }$  and $U(1)_{\rm A}^{\rm F }$ in this model.   In general,  an  axial symmetry $U(1)_A$ in a QCD-like gauge theory is broken by instantons, resulting in a $2K$-fermion operator with a non-vanishing VEV and   $2K$ conserved quantum number \cite{Harari:1981bs}.  Thus we have,
\begin{equation}
 U(1)_A \rightarrow  \mathcal{Z}_{2K},
\end{equation}
where $K$ shows the  massless flavours of the  gauge dynamics in the $N$-dimensional representation of the gauge group $SU(\rm N)$.   This provides the existence of a generic   $\mathcal{Z}_{\rm N} \times \mathcal{Z}_{\rm M} \times \mathcal{Z}_{\rm P}$ flavour symmetry, where $\rm N= 2 \rm K_{\rm TC}$, $\rm M= 2 \rm K_{\rm DTC}$, and $\rm P= 2 \rm K_{\rm F}$.  After the breaking, we have certain conserved axial charges modulo $2K$.  For achieving the  SHVM, as discussed in reference \cite{Abbas:2020frs}, we can use multi-fermion chiral condensates to form the singlet scalar fields $\chi_r$.  These condensates or composite operators contain global axial charge $X_{DTC}$, which is conserved modulo $2K$ \cite{Harari:1981bs}.

\begin{figure}[h]
	\centering
 \includegraphics[width=\linewidth]{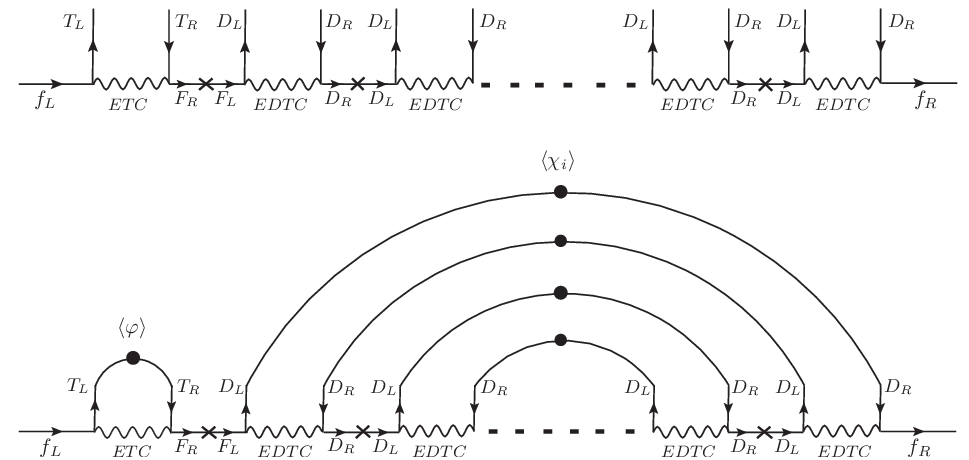}
    \caption{The Feynman diagrams for the masses of the quarks and charged leptons in the  dark-technicolour paradigm.  On the top, the generic interactions of the SM, TC, F and DTC fermions are shown.  In the bottom, we see the formations of a TC on shell chiral condensates, $\langle  \varphi \rangle$ (circular blob), and a   generic   on shell multifermion  chiral condensates  $\langle  \chi_r \rangle$(collection of circular blobs),    and the resulting mass of the SM fermions.}
 \label{fig1}	
 \end{figure}

The masses of charged fermions can be realized through the interactions shown in the upper part of figure  \ref{fig1}.  In the lower part of figure  \ref{fig1}, we show the formation of on-shell TC chiral condensate $\langle  \varphi \rangle$  and  on-shell DTC multifermion chiral condensates   $\langle  \chi_r \rangle$  responsible for providing masses to the charged SM fermions through  the $\mathcal{Z}_{\rm N} \times \mathcal{Z}_{\rm M} \times \mathcal{Z}_{\rm P}$ flavour symmetry.

We notice that the eletroweak scale can be generated by a strong conformal dynamics as recently shown in reference \cite{Chatterjee:2024dgw}. Therefore, we choose the $ SU(\rm N_{\rm TC})$, which produces the eletroweak symmetry breaking via the Higgs mechanism, to be a conformal strong dynamics.  The other two dynamics, the DTC symmetry $ SU(\rm N_{\rm DTC})$ and the dark-QCD $SU(\rm{N}_{\rm F})$ are assumed to be QCD-like theories.

Now, a multi-fermion condensate can be written as   \cite{Aoki:1983yy}, 
\be 
\label{VEV_h}
\langle  ( \bar{\psi}_L \psi_R )^n \rangle \sim \left(  \Lambda \exp(k \Delta \chi) \right)^{3n},
\ee
where $\Delta \chi$ is  the chirality of an operator, $k$ stands for a constant, and $\Lambda$ denotes the scale of the underlying gauge dynamics.   

For a large anomalous dimension, the condensates are given by \cite{Miransky:1994vk},
\begin{align}
\label{TC_con}
\langle \bar{T} T \rangle_{\rm \Lambda_{\rm ETC}}   \approx &  - \dfrac{N_{TC}}{2 \pi^2 \gamma_m} (\Lambda_{TC})^{3-\gamma_m} (\Lambda_{ETC})^{\gamma_m}.
\end{align}
This condensate will be used for the  conformal $ SU(\rm N_{\rm TC})$ symmetry, which provides the electroweak symmetry breaking.

For the DTC symmetry $ SU(\rm N_{\rm DTC})$ and for the dark-QCD $SU(\rm{N}_{\rm F})$, the condensates are given by \cite{Miransky:1994vk},
\begin{align}
\label{DTC_con}
\langle \bar{D} D \rangle_{\rm \Lambda_{\rm DETC}}   \approx &  - \dfrac{N_{DTC}}{4 \pi^2 } \Lambda_{DTC}^{3} \exp(k \Delta \chi),\\ \nonumber
\langle \bar{F} F \rangle_{\rm \Lambda_{\rm GUT}}   \approx &  - \dfrac{N_{F}}{4 \pi^2 } \Lambda^{3} \exp(k_F \Delta \chi).
\end{align}
The mass-matrices in equation \ref{mUD}  of the charged fermions can be approximately generated by the  operators,
\bea
\label{TC_masses1}
\M_{\U,\D,\ell} & = & y_{ij}^f \left[- \frac{g_{\rm ETC}^2}{\Lambda_{\rm ETC}^2} \langle \bar{T} T \rangle_{\rm \Lambda_{\rm ETC}} \right]     \dfrac{1}{\Lambda} \left[- \frac{g_{\rm DETC}^{2n}}{\Lambda_{\rm DETC}^{3n-1}} \left(\langle \bar{D} D \rangle_{\rm \Lambda_{\rm DETC}} \right)^n \right].
\eea
where $n = 1,2,3 \cdots$ and $f=u,d,\ell$.  We can further write the mass matrices in the following form using equations \ref{VEV_h}-\ref{DTC_con},
\bea
\label{TC_masses2}
\M_{\U,\D,\ell} & = & y_{ij}^f \dfrac{N_{TC}}{2 \pi^2 }  \frac{\Lambda_{\text{TC}}^{2}}{\Lambda_{\text{ETC}}}  \dfrac{1}{\Lambda} \left[\dfrac{N_{DTC}}{4 \pi^2 }\right]^{n_i} \frac{\Lambda_{\text{DTC}}^{n_i + 1}}{\Lambda_{\text{EDTC}}^{n_i}} \left[\exp(n_i k) \right]^{n_i/2},~
\eea
where we have assumed $\gamma_m=1$, $g_{\rm ETC} =1$, $g_{\rm DETC}=1$, and  $n_i = 2,4,6, \cdots 2 n $ shows  the number of fermions in a multi-fermion chiral condensate that acts like  the VEV $ \langle \chi_r \rangle$ \cite{Abbas:2020frs}, and $\Lambda_{\text{TC}}$, $\Lambda_{\text{DTC}}$ and  $\Lambda $ stand for  the scale of the TC, DTC, and F dynamics, respectively. 

Using equation \ref{VEV_h},  we  observe the following from  equation \ref{TC_masses2},
\bea
\label{map1}
\epsilon_r \propto \dfrac{1}{\Lambda} \left[\dfrac{N_{DTC}}{4 \pi^2 }\right]^{n_i} \frac{\Lambda_{\text{DTC}}^{n_i + 1}}{\Lambda_{\text{EDTC}}^{n_i}} \left[\exp(n_i k) \right]^{n_i/2}.
\eea
Thus, we can construct the masses of charged fermions given in equation \ref{mass_charged} in terms of equation \ref{map1}, and at the leading order the mass of a charged fermion is given by,
\bea
\label{TC_masses3}
m_{f} & \approx & |y_{11}^f| \dfrac{N_{TC}}{2 \pi^2 }  \frac{\Lambda_{\text{TC}}^{2}}{\Lambda_{\text{ETC}}}  \dfrac{1}{\Lambda} \left[\dfrac{N_{DTC}}{4 \pi^2 }\right]^{n_i} \frac{\Lambda_{\text{DTC}}^{n_i + 1}}{\Lambda_{\text{EDTC}}^{n_i}} \left[\exp(n_i k) \right]^{n_i/2}.
\eea

\begin{figure}[H]
	\centering
 \includegraphics[width=\linewidth]{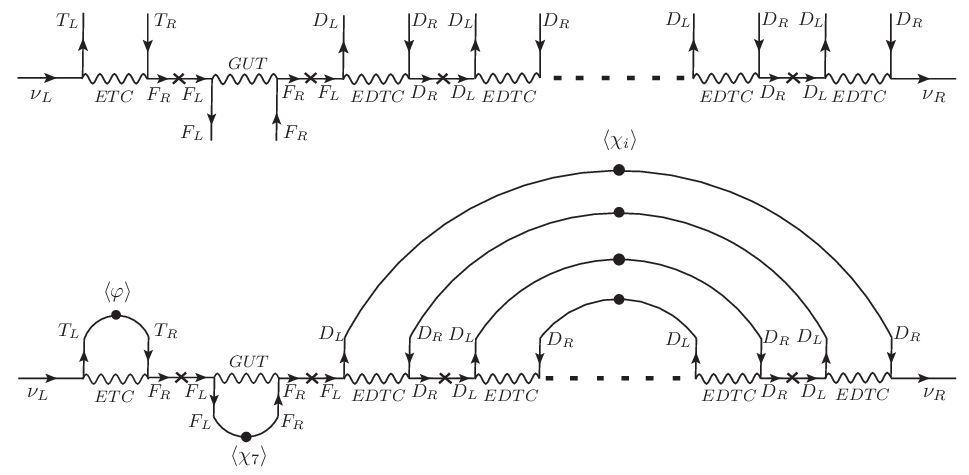}
    \caption{The Feynman diagrams for the masses of neutrinos in the dark-technicolour paradigm. On the top, there are generic interactions involving the SM, TC,  F and DTC gauge sectors mediated by ETC, EDTC and GUT gauge bosons.  In the bottom, we show the generic Feynman diagram after the formation of the fermionic condensates. }
 \label{fig_nu}	
 \end{figure}

The neutrino masses and mixing are achieved by assuming that the ETC and EDTC symmetries eventually can be accommodated in a GUT theory providing  the necessary  dimension-6 operators responsible for neutrino masses given in equation \ref{mass_N}. We show  the required interactions in the upper part  of figure \ref{fig_nu} mediated by the GUT gauge bosons  between the $F_L$ and $F_R$ fermions.   The role of the VEV $\langle \chi_7 \rangle$ is played by the chiral condensate  $\langle \bar{F}_L F_R \rangle$.  The formation of the dimension-6 operators responsible for neutrino masses is shown in the lower part of figure \ref{fig_nu}, where the $\langle  \chi_7 \rangle$ denotes the on-shell chiral condensate (circular blob).

The neutrino mass matrix given in equation \ref{NM} is  recovered as,
\bea
\label{TC_nmassesN}
\M_{\N} & = & y_{ij}^\nu  \dfrac{N_{TC}}{2 \pi^2 }  \frac{\Lambda_{\text{TC}}^{2}}{\Lambda_{\text{ETC}}}  \dfrac{1}{\Lambda} \left[\dfrac{N_{DTC}}{4 \pi^2 }\right]^{n_i} \frac{\Lambda_{\text{DTC}}^{n_i + 1}}{\Lambda_{\text{EDTC}}^{n_i}} \left[\exp(n_i k) \right]^{n_i/2} \dfrac{1}{\Lambda} \dfrac{N_{F}}{4 \pi^2 } \frac{\Lambda^{3}}{\Lambda_{\text{GUT}}^{2}} \exp(2 k_F),
\eea
where,  
\bea
\epsilon_7 \propto  \dfrac{1}{\Lambda} \frac{\Lambda^{3}}{\Lambda_{\text{GUT}}^{2}} \exp(2 k_F).
\eea

The  masses of neutrinos at the leading order are given by,
\bea
\label{neutrino_mass}
m_{\nu} & = & |y_{11}^\nu|   \dfrac{N_{TC}}{2 \pi^2 }  \frac{\Lambda_{\text{TC}}^{2}}{\Lambda_{\text{ETC}}}  \dfrac{1}{\Lambda} \left[\dfrac{N_{DTC}}{4 \pi^2 }\right]^{n_i} \frac{\Lambda_{\text{DTC}}^{n_i + 1}}{\Lambda_{\text{EDTC}}^{n_i}} \left[\exp(n_i k) \right]^{n_i/2} \dfrac{1}{\Lambda} \dfrac{N_{F}}{4 \pi^2 } \frac{\Lambda^{3}}{\Lambda_{\text{GUT}}^{2}} \exp(2 k_F).
\eea

We fit the fermion masses  by mapping the fermionic mass matrices, given in equations \ref{mUD} and equation \ref{NM}, onto equation \ref{TC_masses2} and equation \ref{TC_nmassesN},respectively, and using the  values of the fermion masses at $ 1$TeV  given in reference \cite{Xing:2007fb},
\begin{eqnarray}
\{m_t, m_c, m_u\} &\simeq& \{150.7 \pm 3.4,~ 0.532^{+0.074}_{-0.073},~ (1.10^{+0.43}_{-0.37}) \times 10^{-3}\}~{\rm GeV}, \nonumber \\
\{m_b, m_s, m_d\} &\simeq& \{2.43\pm 0.08,~ 4.7^{+1.4}_{-1.3} \times 10^{-2},~ 2.50^{+1.08}_{-1.03} \times 10^{-3}\}~{\rm GeV},
\nonumber \\
\{m_\tau, m_\mu, m_e\} &\simeq& \{1.78\pm 0.2,~ 0.105^{+9.4 \times 10^{-9}}_{-9.3 \times 10^{-9}},~ 4.96\pm 0.00000043 \times 10^{-4}\}~{\rm GeV}.
\end{eqnarray}

The magnitudes and phases  of the CKM mixing elements are \cite{Zyla:2021},
\bea
|V_{ud}| &=& 0.97370 \pm 0.00014,  |V_{cb}| = 0.0410 \pm 0.0014, |V_{ub}| = 0.00382 \pm 0.00024, \delta = 1.196^{+0.045}_{-0.043}.
\eea

We use the values of the mass square parameters from the global fit  for the normal mass ordering \cite{deSalas:2020pgw},
\bea
\Delta m_{21}^2 &=& (7.50^{+0.64}_{-0.56}) \times 10^{-5} {\rm eV}^2, ~|\Delta m_{31}^2| = (2.55\pm 0.08) \times 10^{-3} \rm{eV}^2,  \\ \nonumber
 \sin \theta_{12}^\ell    &=&  (0.564_{-0.043}^{+0.044}),~
\sin  \theta_{23}^\ell  =   (0.758_{-0.099}^{+0.023}),~   \sin \theta_{13}^\ell  =  (0.1483_{-0.0069}^{+0.0067}),
\eea
where the errors are in 3 $\sigma$ range.

We fit quark masses and mixing  by defining the  $\chi^2_q$ as,
\bea
\chi^2_q &=& \dfrac{(m_q - m_q^{\rm{model}} )^2}{\sigma_{m_q}^2}  + \dfrac{(\sin \theta_{ij} - \sin \theta_{ij}^{\rm{model}} )^2}{\sigma_{\sin \theta_{ij}}^2} ,
\eea
where $q=\{u,d,c,s,t,b\}$ and $i,j=1,2,3$.

We scan the dimensionless coefficients $y_{ij}^{u,d}= |y_{ij}^{u,d}| e^{i \phi_{ij}^{q}}$  in the range $|y_{ij}^{u,d}| \in [0.3, 4 \pi]$ and $ \phi_{ij}^{q} \in [0,2\pi]$. The result of fitting is
 \begin{eqnarray}
 \label{fit}
 && \rm{N}_{\rm TC} = 12, \rm{N}_{\rm DTC} = 40, 
    \Lambda_{\rm TC}= 381  \text{ GeV} , \Lambda_{\rm ETC}= 10^7 \text{ GeV}, \Lambda_{\rm DTC}= 10^3 \text{ GeV},   
 \Lambda_{\rm EDTC}= 1.26 \times 10^3 \text{ GeV}, \nonumber \\
&& \Lambda = 2 \times 10^3 \text{ GeV},  
   \{n_1, n_2, n_3, n_4, n_5, n_6\}  = \{2, 4, 6, 2, 2, 4\}, k=0.2, \delta = 1.196.
 \end{eqnarray}
The dimensionless couplings $y_{ij}^{u,d}$ are,
 \begin{equation}
y^u_{ij} = \begin{pmatrix}
0.005\, -0.3 i & 0 & 0.54\, +0.47 i  \\
0 & 1.44\, -2.54 i &  9.92\, +6.69 i \\
0 & 3.41 & 1.35\, -2.3 i
\end{pmatrix},  
\end{equation}
\begin{equation}
y^d_{ij} = \begin{pmatrix}
-0.25+0.46 i & -0.97-1.61 i & -0.74+0.14 i   \\
-0.20+0.46 i & -8.28-0.09 i & -10.13-1.0 i \\
0.21\, +0.46 i & -0.35-8.52 i & 6.75\, -8.56 i
\end{pmatrix},  
\end{equation}
These results are obtained for $\chi^2_{q,min} = 22.18$\footnote{A more sophisticated fit  will be performed using  the renormalization-group evolution and the Dyson-Schwinger equations of the model in a future study.}.

Using the fit results of the quark sector, we fit the masses of charged leptons, and parameters of neutrino masses and mixing by defining,
\bea
\chi^2_N &=&   \dfrac{(m_\ell - m_\ell^{\rm{model}} )^2}{\sigma_{m_\ell}^2} + \dfrac{(\Delta m_{21}^2 - \Delta m_{21}^{2,\rm{model}} )^2}{\sigma_{\Delta m_{21}^2}^2}  +  \dfrac{(\Delta m_{31}^2 - \Delta m_{31}^{2,\rm{model}} )^2}{\sigma_{\Delta m_{31}^2}^2} +  \dfrac{(\sin \theta_{ij}^\ell - \sin \theta_{ij}^{\ell\rm{model}} )^2}{\sigma_{\sin \theta_{ij}^\ell}^2},
\eea
where  $\ell=\{e,\mu,\tau\}$, and $i,j=1,2,3$. We obtain
\begin{eqnarray}
 \label{fit}
 &&  \rm{N}_{\rm F} = 3,  \Lambda_{\rm GUT}= 2 \times 10^{16} \text{ GeV}, \Lambda = 2 \times 10^3 \text{ GeV},n_7=2,  k_F = 7.64.
 \end{eqnarray}

The dimensionless couplings $y_{ij}^{\ell,\nu}$ are,
\begin{equation}
y^\ell_{ij} = \begin{pmatrix}
0.41\, -0.91 i & 1 & 1 \\
0  & -0.93+0.59 i & 1 \\
0 & 0 & 0.19\, 1.1 i
\end{pmatrix},
\end{equation}
\begin{equation}
y^\nu_{ij} = \begin{pmatrix}
0.93\, +0.25 i & 1 & 1 \\
0 & 1.16\, -0.24 i & -0.84-0.88 i \\
0 & 1.7\, -0.3 i & 0.034\, -1.48 i
\end{pmatrix}, 
\end{equation}
These results are obtained for $\chi^2_{N,min} = 3$.

\subsection{UV completion of the FN mechanism}
\begin{figure}[H]
	\centering
 \includegraphics[width=\linewidth]{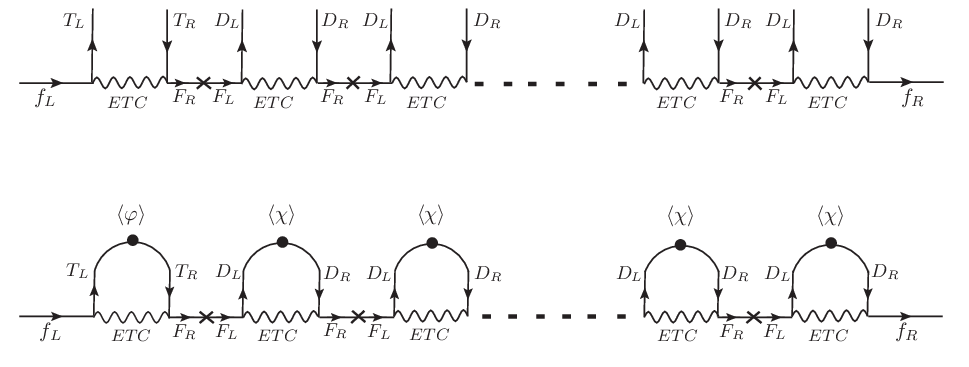}
    \caption{Feynman diagrams for the masses of the quarks and charged leptons in the  dark-technicolour paradigm.  On the top, the generic interactions of the SM, TC, F and DTC fermions are shown.  In the bottom, we see the formations of the  condensate (circular blob),    and the resulting mass of the SM fermions.}
 \label{fig_DTC_FN}	
 \end{figure}

For achieving the UV completion of the FN mechanism in the dark-technicolour paradigm, we need to assume that the SM, TC, and DTC dynamics are accommodated in a common extended technicolour  symmetry.  This condition is not necessary for the UV completion of the SHVM, and provides a crucial difference between the SHVM and the FN mechanism in the dark-technicolour paradigm.  This leads to the  required interactions for the charged fermions masses  in the FN mechanism, which are shown in the upper part of figure  \ref{fig_DTC_FN}.  The mass generation through the FN mechanism is shown in the lower part of figure  \ref{fig_DTC_FN}, where  $\langle \varphi \rangle$ and $\langle \chi \rangle$  are the  on-shell chiral condensate acting like the VEVs of the Higgs and flavon fields.

Now, the masses of the SM fermions in the FN mechanism originate as,
\bea
\label{TC_masses}
m_{f} & \approx &   |y_{ii}^f| \dfrac{N_{TC}}{2 \pi^2 }  \frac{\Lambda_{\text{TC}}^{2}}{\Lambda_{\text{ETC}}}  \left( \dfrac{1}{\Lambda} \dfrac{N_{DTC}}{4 \pi^2 }\frac{\Lambda_{\text{DTC}}^3}{\Lambda_{\text{ETC}}^{2}} \exp(2 k) \right)^{n_{ij}^f},
\eea
where $f=u,d$, and the  order parameter $\epsilon$  can be identified as, 
\begin{align}
    \epsilon  \propto \dfrac{1}{\Lambda} \dfrac{N_{DTC}}{4 \pi^2 }\frac{\Lambda_{\text{DTC}}^3}{\Lambda_{\text{ETC}}^{2}} \exp(2 k),  
\end{align}
and the SM Higgs VEV can be written as,
\begin{align}
  \langle \varphi \rangle \propto   \dfrac{N_{TC}}{2 \pi^2 }  \frac{\Lambda_{\text{TC}}^{2}}{\Lambda_{\text{ETC}}}. 
\end{align}

\section{Beyond the dark-technicolour paradigm: An extended  and dark-extended technicolour}
\label{ETC}
In this section, we sketch an outline of an extended  and dark-extended technicolour scenario for the dark-technicolour paradigm, which leads to the SHVM at low energies.  We define the ETC fermionic sector in an extended $SU(\rm N_{\rm TC} + 12 ) $ as follows:
\begin{eqnarray}
\psi^{\rm ETC}_L  &\equiv&   \begin{pmatrix}
T ~B  \\
\cdots \\
\cdots \\
T ~B  \\
u ~d \\
u ~d \\
u ~d \\
\nu_e~ e\\
c ~s \\
c ~s \\
c ~s \\
\nu_\mu~ \mu\\
t ~b \\
t ~b \\
t ~b \\
\nu_\tau~ \tau
\end{pmatrix}_L,     
\psi^{\rm ETC}_R  \equiv   \begin{pmatrix}
T  \\
\cdots \\
\cdots \\
T  \\
U^1 \\
U^1  \\
U^1  \\
N^1\\
U^2 \\
U^2  \\
U^2  \\
N^2\\
U^3 \\
U^3  \\
U^3  \\
N^3\\
\end{pmatrix}_R,   
\psi^{\rm ETC}_R  \equiv   \begin{pmatrix}
B  \\
\cdots \\
\cdots \\
B  \\
D^1 \\
D^1  \\
D^1  \\
E^1\\
D^2 \\
D^2  \\
D^2  \\
E^2\\
D^3 \\
D^3  \\
D^3  \\
E^3\\
\end{pmatrix}_R,    
\end{eqnarray}

The dark-ETC is defined by the group $SU(\rm N_{\rm DTC} +1)$ where the first family quark-multiplet takes the following form:

\begin{eqnarray}
\psi^{\rm DETC,q}_{L,i}  &\equiv&   \begin{pmatrix}
c_i ~c_i~c_i~c_i \cdots U^i \\
s_i ~s_i~s_i~s_i \cdots D^i \\
\end{pmatrix}_L,     
\psi^{\rm DETC, q}_R  \equiv  \begin{pmatrix}
c_i ~c_i~c_i~c_i \cdots u_i \\
s_i ~s_i~s_i~s_i \cdots d_i \\
\end{pmatrix}_R,   
\end{eqnarray}
where $i=1,2,3 \cdots$ is a generation label, and $u_i= u, c, t$, and $d_i= d, s, b$ are the right-handed quark SM fields.

In a similar way, the leptonic multiplet can be written as,
\begin{eqnarray}
\psi^{\rm DETC,\ell}_{L,i}  &\equiv&   \begin{pmatrix}
e_i ~e_i~e_i~e_i \cdots N^i \\
n_i ~n_i~n_i~n_i \cdots E^i \\
\end{pmatrix}_L,     
\psi^{\rm DETC,\ell}_{R,i}  \equiv   \begin{pmatrix}
e_i ~e_i~e_i~e_i \cdots \nu_i \\
n_i ~n_i~n_i~n_i \cdots e_i \\
\end{pmatrix}_R,      
\end{eqnarray}
where $\nu_i= \nu_e,  \nu_\mu, \nu_\tau$ and $e_i= e, \mu, \tau$ are the right-handed leptonic SM fields.

We assume that there exists a GUT symmetry $\mathcal{G}_U$, which is broken along the following track,
\be 
\mathcal{G}_U \rightarrow SU(\rm N_{\rm TC} + 12 ) \times SU(\rm N_{\rm F}) \times SU(\rm N_{\rm DTC} + 1 ) \rightarrow  SU(\rm N_{\rm TC} )  \times SU(\rm N_{\rm F}) \times SU(\rm N_{\rm DTC}).
\ee

\section{Summary}
\label{sec7}
The flavour problem of the SM is a  subtle and fascinating puzzle whose solution may lead to the physics beyond the SM. The earliest systematic attempt to solve this problem was made by Weinberg by producing light fermion masses through radiative corrections.  This was followed by various approaches, such as models based on continuous and discrete symmetries, GUT theories, extra dimensions, and strong dynamics, such as theories based on TC scenarios.

In this review, we have particularly focused on models based on discrete symmetries, which have possibilities to provide a unified solutions to flavour and dark matter.  We notice that although discrete symmetries are extensively used in neutrino model building, not much efforts are being made to solve the flavour problem through, in particular, abelian discrete symmetries.  In addition to reviewing the existing approaches to the flavour problem, we have presented a thorough discussion of two new solutions of the flavour problem,  which are based on the $\mathcal{Z}_N \times \mathcal{Z}_M$ flavour symmetry and the SHVM framework.  We have shown that both of these approaches may have a common UV origin based on a dark technicolour paradigm, where the flavon field and the gauge singlet scalar fields are realized in the form of the chiral DTC condensates.  Moreover, the $\mathcal{Z}_N \times \mathcal{Z}_M \times \mathcal{Z}_P$ flavour symmetry can also provide the so-called flavonic dark matter together with a solution to the flavour problem.  Furthermore, the SHVM also contains a possible dark matter candidate, which is the axial degree of freedom of the field $\chi_7$.

We notice that in the light of emerging hints  of new physics in different experiments, the new models discussed in this work may play an important role in explaining the origin of these anomalies.  For instance, we observe that the scalar mass spectrum of the SHVM is hierarchical due to the existence of the hierarchical VEVs.  Therefore, the SHVM may be a useful framework to study the anomalies, given in table \ref{tab:anomalies} by the ATLAS experiments.   Furthermore, we have the dark-technicolour paradigm based on the $ SU(\rm N_{\rm TC}) \times SU(\rm N_{\rm DTC}) \times SU(\rm{N}_{\rm F})$ symmetry symmetry.  The SHVM and the FN mechanism may be a low-energy effective limit of this framework.

\section*{Acknowledgments}
This  work is supported by the  Council of Science and Technology,  Govt. of Uttar Pradesh,  India through the  project ``   A new paradigm for flavour problem "  no.   CST/D-1301, and Anusandhan National Research Foundation (Science and Engineering Research Board) , Department of Science and Technology, Government of India through the project `` Higgs Physics within and beyond the Standard Model" no. CRG/2022/003237. NS acknowledges the support through the INSPIRE fellowship by the Department of Science and Technology, Government of India.

\section*{Data availability statements}
The used data is explicitly quoted in the manuscript itself, and there is no need to deposit it separately.





\end{document}